\documentclass[10pt, final, twocolumn, twoside, romanappendices]{IEEEtran}
\usepackage{amsmath,amsfonts}
\usepackage{amssymb}
\usepackage{algorithmic}
\usepackage{algorithm}
\usepackage{array}
\usepackage[caption=false,font=footnotesize,labelfont=sf,textfont=sf]{subfig}
\usepackage{textcomp}
\usepackage{stfloats}
\usepackage{url}
\usepackage{verbatim}
\usepackage{graphicx}
\usepackage{cite}
\usepackage{color}
\usepackage{acronym}
\usepackage{units}

\usepackage{soul}
\usepackage{float} 

\begin{document}

\bstctlcite{IEEEexample:BSTcontrol}

\newcommand{\bluenote}[1]{{\color{blue}$\blacktriangleright${#1}$\blacktriangleleft$}}
\newcommand{\nic}[1] {{\color{blue}{#1}}}
\newcommand{\gp}[1] {{\color{red}{#1}}}

%
\acrodef{AoA}{angle-of-arrival}

\acrodef{AoD}{angle-of-departure}

\acrodef{DoA}{direction of arrival}

\acrodef{ID}{identifier}

\acrodef{V2X}{vehicle-to-everything}

\acrodef{IIoT}{industrial Internet-of-things}

\acrodef{RAA}{retro-directive antenna array}

\acrodef{IoT}{Internet-of-things}

\acrodef{AP}{access point}

\acrodef{ABIM}{active backscattering intelligent metasurface}

\acrodef{SVD}{singular-value decomposition}

\acrodef{EVD}{eigen-value decomposition}

\acrodef{PSWF}{prolate spheroidal wave function}

\acrodef{CR}{channel response}

\acrodef{BS}{base station}

\acrodef{MS}{mobile station}

\acrodef{UE}{user equipment}

\acrodef{MIMO}{multiple-input multiple-output}

\acrodef{MU-MIMO}{multi-user MIMO}

\acrodef{mMTC}{massive Machine Type Communication}

\acrodef{RIS}{reconfigurable intelligent surface}

\acrodef{IRS}{intelligent reconfigurable surface}

\acrodef{LIS}{large intelligent surface}

\acrodef{MIS}{medium intelligent surface}

\acrodef{SIS}{small intelligent surface}

\acrodef{DoF}{degrees-of-freedom}

\acrodef{AF}{amplify \& forward}

\acrodef{DF}{detect \& forward}

\acrodef{JF}{just forward}

\acrodef{CSI}{channel state information}

\acrodef{RV}{random variable}

\acrodef{i.i.d.}{independent, identically distributed}

\acrodef{PSD}{power spectral density}

\acrodef{PDF}{probability distribution function}

\acrodef{CDF}{cumulative distribution function}

\acrodef{ch.f.}{characteristic function}

\acrodef{AWGN}{additive white Gaussian noise}

\acrodef{RSSI}{received signal strength indicator}

\acrodef{SNR}{signal-to-noise ratio}

\acrodef{SINR}{signal-to-interference-noise ratio}

\acrodef{LRT}{likelihood ratio test}

\acrodef{GLRT}{generalized likelihood ratio test}

\acrodef{GML}{generalized maximum likelihood}

\acrodef{LOS}{line-of-sight}

\acrodef{NLOS}{non-line-of-sight}

\acrodef{GDOP}{geometric dilution of precision}

\acrodef{GPS}{Global Positioning System}

\acrodef{FIM}{Fisher information matrix}

\acrodef{PEB}{position error bound}

\acrodef{WSN}{Wireless Sensor Network}

\acrodef{MAC}{medium access control}

\acrodef{RSS}{received signal strength}

\acrodef{RTT}{round-trip time}

\acrodef{MIMO}{multiple-input multiple-output}

\acrodef{MF}{matched filter}

\acrodef{ED}{energy detector}

\acrodef{ML}{maximum likelihood}

\acrodef{NL}{nonlinear}

\acrodef{MSE}{mean square error}

\acrodef{RMSE}{root mean square error}

\acrodef{ppm}{part-per-million}

\acrodef{PRP}{pulse repetition period}

\acrodef{ACK}{acknowledge}

\acrodef{UWB}{ultrawide bandwidth}

\acrodef{TNR}{threshold-to-noise ratio}

\acrodef{NLOS}{non line-of-sight}

\acrodef{LOS}{line-of-sight}

\acrodef{LS}{least squares}

\acrodef{IR-UWB}{impulse radio UWB}

\acrodef{FCC}{Federal Communications Commission}
\acrodef{node A}{Node A}
\acrodef{node B}{Node B}
\acrodef{TH}{time-hopping}

\acrodef{PPM}{pulse position modulation}

\acrodef{PAM}{pulse amplitude modulation}

\acrodef{MUI}{multi-user interference}

\acrodef{PDP}{power delay profile}

\acrodef{PPP}{Poisson point process}

\acrodef{DS}{delay spread}

\acrodef{CED}{channel excess delay}

\acrodef{BPZF}{band-pass zonal filter}

\acrodef{SIR}{signal-to-interference ratio}

\acrodef{RFID}{radio frequency identification}

\acrodef{WPAN}{wireless personal area networks}

\acrodef{WWLB}{Weiss-Weinstein lower bound}

\acrodef{DP}{direct path}

\acrodef{MF}{matched filter}

\acrodef{MMSE}{minimum-mean-square-error}

\acrodef{SBS}{serial backward search}

\acrodef{NBI}{narrowband interference}

\acrodef{WBI}{wideband interference}

\acrodef{INR}{interference-to-noise ratio}

\acrodef{CIR}{channel impulse response}

\acrodef{ISI}{inter-symbol interference}

\acrodef{CPR}{channel pulse response}

\acrodef{LRT}{likelihood ratio test}

\acrodef{RADAR}{RADAR}

\acrodef{MUR}{Multistatic RADAR}

\acrodef{MUI}{multi-user interference}

\acrodef{EM}{electromagnetic}

\acrodef{CW}{continuous wave}

\acrodef{RF}{radiofrequency}

\acrodef{FCC}{Federal Communications Commission}

\acrodef{EIRP}{effective radiated isotropic power}

\acrodef{ECDF}{empirical cumulative distribution function}

\acrodef{RCS}{radar cross-section}

\acrodef{BAV}{balanced antipodal Vivaldi}

\acrodef{PRake}{partial Rake}

\acrodef{RTLS}{real time locating system}

\acrodef{CRB}{Cram\'{e}r-Rao bound}

\acrodef{ZZB}{Ziv-Zakai bound}

\acrodef{TOA}{time-of-arrival}

\acrodef{TOF}{time-of-flight}

\acrodef{WSN}{wireless sensor network}

\acrodef{MAC}{medium access control}

\acrodef{RSS}{received signal strength}

\acrodef{TDOA}{time difference-of-arrival}

\acrodef{RF}{radiofrequency}

\acrodef{RTT}{round-trip time}

\acrodef{AOA}{angle-of-arrival}

\acrodef{MF}{matched filter}

\acrodef{ED}{energy detector}

\acrodef{ML}{maximum likelihood}

\acrodef{MUR}{Multistatic radar}

\acrodef{HDSA}{high-definition situation-aware}

\acrodef{RRC}{root raised cosine}

\acrodef{OFDM}{orthogonal frequency division multiplexing}

\acrodef{IF}{intermediate frequency}

\acrodef{PHY}{physical layer}

\acrodef{S-V}{Saleh-Valenzuela}

\acrodef{UHF}{ultra-high frequency}

\acrodef{PR}{pseudo-random}

\acrodef{SoC}{System on Chip}

\acrodef{SoP}{System on Package}

\acrodef{SPMF}{Single-Path Matched Filter}

\acrodef{IMF}{Ideal Matched Filter}

\acrodef{SCR}{signal-to-clutter ratio}

\acrodef{BEP}{bit error probability}

\acrodef{BER}{bit error rate}

\acrodef{WSR}{wireless sensor radar}

\acrodef{HPBW}{half power beam width}

\acrodef{LEO}{localization error outage}

\acrodef{WSS}{wide-sense stationary}

\acrodef{TR}{time-reversal}

\acrodef{MIMO TRX}{MIMO transceiver}

\acrodef{WSSUS}{WSS with uncorrelated scattering}

\acrodef{GP}{Gaussian process}

\acrodef{IMU}{inertial measurement unit}

\acrodef{TDD}{time-division duplexing}

\acrodef{ULA}{uniform linear array}

\acrodef{SCM}{self-conjugate metasurface}

\acrodef{CDL}{Clustered Delay Line}

\acrodef{PN}{pseudo noise}

\acrodef{GNSS}{global navigation satellite system}

\acrodef{UAV}{unmanned aerial vehicle}

\acrodef{SATCOM}{Satellite Communication}
\acrodef{FFT}{Fast Fourier Transform}

\acrodef{DFT}{Discrete Fourier Transform}


\newcommand{\rednote}[1] {{\color{red}$\blacktriangleright${#1}$\blacktriangleleft$}}
\newcommand{\blue}[1] {{\color{blue}{#1}}}
\newcommand{\virg}[1] {``{{#1}}''}

\newcommand{\fig}[1]{Fig.~\ref{#1}}
\newcommand{\sect}[1]{Sec.~\ref{#1}}
\newcommand{\apd}[1]{Appendix~\ref{#1}}
\newcommand{\eq}[1]{(\ref{#1})}

%



\newcommand{\Real}[1]{\Re \left \{ #1\right \}}
\newcommand{\en} {E}
\newcommand{\td}[1] {\tilde{#1}}
\newcommand{\lt}[1] {{\td{\lambda}}_{#1}}
\newcommand{\bl}[1] {\text{\boldmath ${\lambda}$}_{#1}}
\newcommand{\blt}[1] {\text{\boldmath $\td{\lambda}$}_{#1}}
\newcommand{\vg}[1] {{\mbox{{\boldmath ${#1}$}}}}
\newcommand{\vgs}[2] {\vg{#1}_{#2}}

\newcommand{\fourier}[1]{\mathcal{F} \left [ #1\right ]}
\newcommand{\invfourier}[1]{\mathcal{F}^{-1} \left [ #1\right ]}

\newcommand{\PX}[1] {{\mathbb{P}}\left\{{#1}\right\}}
\newcommand{\EX}[1] {{\mathbb{E}}\left\{{#1}\right\}}
\newcommand{\EXs}[2] {{\mathbb{E}}_{{#1}}\!\!\left\{{#2}\right\}}
\newcommand{\Var}[1] {{\text{Var}}\left ({#1}\right )}

\newcommand{\pX}[1] {{\mathbf{p}}\left\{{#1}\right\}}

\newcommand{\Hone} {\mathcal{H}_1}
\newcommand{\Hzero}{\mathcal{H}_0}
\newcommand{\hatHone} {\hat{\mathcal{H}}_1}
\newcommand{\hatHzero}{\hat{\mathcal{H}}_0}

\newcommand{\Mu} {\mathcal{M}_u}
\newcommand{\Mc}{\mathcal{M}_c}

\newcommand{\boldA} {{\bf{A}}}
\newcommand{\boldg} {{\bf{g}}}
\newcommand{\bolds} {{\bf{s}}}
\newcommand{\boldf} {{\bf{f}}}
\newcommand{\bolda} {{\bf{a}}}
\newcommand{\boldb} {{\bf{b}}}
\newcommand{\boldp} {{\bf{p}}}
\newcommand{\bolde} {{\bf{e}}}
\newcommand{\boldk} {{\bf{k}}}
\newcommand{\boldK} {{\bf{K}}}
\newcommand{\boldu} {{\bf{u}}}
\newcommand{\boldc} {{\bf{c}}}
\newcommand{\boldV} {{\bf{V}}}
\newcommand{\boldX} {{\bf{X}}}
\newcommand{\boldY} {{\bf{Y}}}
\newcommand{\boldW} {{\bf{W}}}
\newcommand{\boldU} {{\bf{U}}}
\newcommand{\boldE} {{\bf{E}}}
\newcommand{\boldJ} {{\bf{J}}}
\newcommand{\boldH} {{\bf{H}}}
\newcommand{\boldm} {{\bf{m}}}
\newcommand{\boldP} {{\bf{P}}}
\newcommand{\boldF} {{\bf{F}}}
\newcommand{\boldG} {{\bf{G}}}
\newcommand{\boldR} {{\bf{R}}}
\newcommand{\boldC} {{\bf{C}}}
\newcommand{\boldB} {{\bf{B}}}
\newcommand{\boldD} {{\bf{D}}}
\newcommand{\boldSigma} {{{\Sigma}}}

\newcommand{\boldLambda} {{\bf{\Lambda}}}
\newcommand{\boldq} {{\bf{q}}}

\newcommand{\boldI} {{\bf{I}}}
\newcommand{\boldr} {{\bf{r}}}
\newcommand{\meanr} {{\overline{r}}}
\newcommand{\meanboldr} {{\overline{\bf{r}}}}
\newcommand{\boldn} {{\bf{n}}}
\newcommand{\boldx} {{\bf{x}}}
\newcommand{\boldy} {{\bf{y}}}
\newcommand{\boldh} {{\bf{h}}}
\newcommand{\boldz} {{\bf{z}}}
\newcommand{\boldw} {{\bf{w}}}
\newcommand{\boldv} {{\bf{v}}}

\newcommand{\boldt} {{\bf{t}}}
\newcommand{\meanw} {{\overline{w}}}
\newcommand{\meanboldw} {{\overline{\bf{w}}}}
\newcommand{\boldd} {{\bf{d}}}
\newcommand{\boldalpha} {\bf{\alpha}}
\newcommand{\boldbeta} {\bf{\beta}}
\newcommand{\boldgamma} {\bf{\gamma}}
\newcommand{\boldrho} {\bf{\rho}}
\newcommand{\boldhc} {\bf{h}_{\text{c}}}

\newcommand{\Pd} {P_{\text{d}}}
\newcommand{\Pf} {P_{\text{f}}}

\newcommand{\Pb} {P_{\text{b}}}

\newcommand{\Rb} {R_{\text{b}}}
\newcommand{\Ep} {E_{\text{p}}}

\newcommand{\Tp} {T_{\text{p}}}
\newcommand{\Td} {T_{\text{d}}}
\newcommand{\fc} {f_{\text{c}}}
\newcommand{\ts} {t_{\text{s}}}
\newcommand{\Ta} {T_{\text{a}}}
\newcommand{\Ti} {T_{\text{i}}}
\newcommand{\Np} {N_{\text{p}}}
\newcommand{\Nps} {N_{\text{ps}}}
\newcommand{\tp} {\tau_{\text{p}}}
\newcommand{\Es} {E_{\text{s}}}
\newcommand{\Eb} {E_{\text{b}}}
\newcommand{\Ts} {T_{\text{s}}}
\newcommand{\Tf} {T_{\text{f}}}
\newcommand{\Tc} {T_{\text{c}}}
\newcommand{\Th} {T_{\text{h}}}
\newcommand{\Tb} {T_{\text{b}}}
\newcommand{\Tob} {T_{\text{ob}}}
\newcommand{\Nc} {N_{\text{c}}}
\newcommand{\Ns} {N_{\text{s}}}
\newcommand{\Na} {N_{\text{A}}}

\newcommand{\Nsym} {N_{\text{sym}}}
\newcommand{\tint} {T_{\text{int}}}
\newcommand{\TX}[1] {{\mathbb{T}}\left [{#1}\right ]}
\newcommand{\Prob}[1] {\text{P}\left\{{#1}\right\}}
\newcommand{\Q}[1] {Q \left ( #1 \right )}
\newcommand{\Nch} {N_{\text{ch}}}
\newcommand{\Lp} {L_{\text{p}}}
\newcommand{\dref} {d_{\text{ref}}}
\newcommand{\wref} {w_{\text{ref}}}
\newcommand{\Wref} {W_{\text{ref}}}
\newcommand{\Href} {H_{\text{ref}}}
\newcommand{\ZA} {Z_{\text{A}}}
\newcommand{\taup} {\tau_{\text{f}}}
\newcommand{\taud} {{\tau_{\text{d}}}}
\newcommand{\etaud} {\hat{\tau}_{\text{d}}}
\newcommand{\toa} {\tau}
\newcommand{\etoa} {\hat{\tau}}
\newcommand{\Beff} {B_{\text{eff}}}

\newcommand{\Pric} {P_{\text{r}}}
\newcommand{\Thetai} {{\bf \Theta}^{\text{inc}}}
\newcommand{\Thetar} {{\bf \Theta}^{\text{ref}}}
\newcommand{\Thetat} {{\bf \Theta}^{\text{t}}}
\newcommand{\Prc}{P_{r}^{metal\,can}}
\newcommand{\Prw}{P_{r}^{bottle\,water}}

\newcommand{\floor}[1] {f \left ({#1} \right )}
\newcommand{\rect}[1] {\text{Rect} \left ({#1} \right )}
\newcommand{\sinc}[1] {\text{sinc} \left ({#1} \right )}

\def\dsp{\displaystyle}

\def\erfc{{\text{erfc}}}
\def\erf{{\text{erf}}}
\def\inve{{\text{inverfc}}}
\def\teq{\triangleq}
\def\bs{$\blacksquare$}
\newcommand{\tr}[1]{{\rm tr} \left ( #1 \right ) }
\newcommand{\rank}[1]{{\rm rank} \left ( #1 \right )}
\newcommand{\diag}[1]{{\rm diag} \left ( #1 \right )}

\newcommand{\Cfunc}[1]{{C^{(\text{#1})}}}
\newcommand{\cvect}[1]{{\mathbf{c}^{(\text{#1})}}}
\newcommand{\ucvect}[1]{{\underline{{\mathbf{c}}}^{(\text{#1})}}}
\newcommand{\epsilonvect}[1]{{\mathbf{\varepsilon}^{(\text{#1})}}}
\newcommand{\uchat}[1]{{\underline{\hat{\mathbf{c}}}^{(\text{#1})}}}
\newcommand{\chat}[1]{{\hat{\mathbf{c}}^{(\text{#1})}}}

\newcommand{\dEve} {d_{\text{Eve}}}
\newcommand{\NEve} {N_{\text{Eve}}}
\newcommand{\MEve} {M_{\text{Eve}}}

\newcommand{\Dset} {\mathcal{D}}
\newcommand{\GP}[1] {\mathcal{GP}\left ( #1 \right )}

\newcommand{\argmax}[1]{\underset{{#1}}{\operatorname{argmax}}}
\newcommand{\argmin}[1]{\underset{{#1}}{\operatorname{argmin}}}

\newcommand{\convZ}{*}
\newcommand{\conjZ}{^{\dag}}
\newcommand{\argmaxZ}[1]{\operatorname*{argmax}_{#1}}
\newcommand{\argminZ}[1]{\operatorname*{argmin}_{#1}}
\newcommand{\sincZ}{\text{sinc}}

\newcommand{\EXZ}[1] {{\mathbb{E}}\left\{{#1}\right\}}
\newcommand{\EXZBig} {\mathbb{E}}

\newcommand{\0}{\mathbf{0}}


\newcommand{\RZ}{\mathbb{R}^2}
\newcommand{\A}{\mathcal{A}}
\newcommand{\Nset}{\mathcal{N}}
\newcommand{\Vset}{\mathcal{V}}

\newcommand{\rc}{r_{\text{c}}}
\newcommand{\Pe}{P_{\text{e}}}
\newcommand{\SNR}{\mathsf{SNR}}

\newcommand{\bx} {{\bf{x}}}
\newcommand{\bX} {{\bf{X}}}
\newcommand{\bW} {{\bf{W}}}
\newcommand{\bw} {{\bf{w}}}
\newcommand{\bY} {{\bf{Y}}}
\newcommand{\boldeta} {{\boldsymbol{\eta}}}

\newcommand{\Nr} {N_{\text{R}}}
\newcommand{\Nt} {N_{\text{T}}}
\newcommand{\Nmin} {N_{\text{min}}}
\newcommand{\Lr} {L_{\text{R}}}
\newcommand{\Lt} {L_{\text{T}}}

\newcommand{\Ptx} {P_{\text{T}}}
\newcommand{\Prx} {P_{\text{R}}}

\newcommand{\Gt} {G_{\text{T}}}
\newcommand{\Gr} {G_{\text{R}}}
\newcommand{\Gi} {G_{\text{I}}}
\newcommand{\Glis} {G_{\text{LIS}}}
\newcommand{\Gap} {G_{\text{A}}}
\newcommand{\Gscm} {G_{\text{RAA}}}
\newcommand{\Fap} {F_{\text{A}}}
\newcommand{\Fscm} {F_{\text{RAA}}}

\newcommand{\SNRmax} {\mathsf{SNR}_{1,\mathrm{max}}}
\newcommand{\SNRboot} {\mathsf{SNR}_{1,\mathrm{boot}}}
\newcommand{\SNRdec} {\mathsf{SNR}_{\mathrm{dec}}}
\newcommand{\SNRjmax} {\mathsf{SNR}_{j,\mathrm{max}}}
\newcommand{\SNRth} {\mathsf{SNR}_{1,\mathrm{th}}}
\newcommand{\SNRstart} {\mathsf{SNR}_{\mathrm{start}}}

\title{ Localization Based on MIMO Backscattering from \\ Retro-Directive Antenna Arrays}

\author{ \IEEEauthorblockN{
 Marina Lotti,~\IEEEmembership{Graduate~Student~Member,~IEEE}, 
 Nicol\`o  Decarli,~\IEEEmembership{Member,~IEEE}, \\
 Gianni Pasolini,~\IEEEmembership{Member,~IEEE}, and
 Davide Dardari,~\IEEEmembership{Senior~Member,~IEEE}   
}

\thanks{Manuscript submitted 22 Apr. 2024.}
\thanks{This work was partially supported by the European Union under the Italian National Recovery and Resilience Plan (NRRP) of NextGenerationEU, partnership on ``Telecommunications of the Future" (PE00000001 - program ``RESTART") and under the National Recovery and Resilience Plan (NRRP), Mission 04 Component 2 Investment 1.5 – NextGenerationEU, Call for tender n. 3277 dated 30/12/2021, Award Number:  0001052 dated 23/06/2022.}
\thanks{M.~Lotti, G.~Pasolini and D.~Dardari are with the Dipartimento di Ingegneria dell'Energia Elettrica e dell'Informazione ``Guglielmo Marconi" (DEI), University of Bologna, and WiLab-CNIT,   Bologna (BO), Italy, e-mail: \{marina.lotti2, gianni.pasolini, davide.dardari\}@unibo.it.\\
N.~Decarli is with the National Research Council - Institute of Electronics, Computer and Telecommunication Engineering (CNR-IEIIT), and  WiLab-CNIT, Bologna (BO), Italy,  e-mail: nicolo.decarli@cnr.it.\\
}}

\markboth{Submitted for review to IEEE Trans. on Vehicular Technologies}%
{Lotti M. \MakeLowercase{\textit{et al.}}: Localization Based on MIMO Backscattering From Retro-Directive Antenna Arrays}

\maketitle

\begin{abstract}
In next-generation vehicular environments, precise localization is crucial for facilitating advanced applications such as autonomous driving. As automation levels escalate, the demand rises for enhanced accuracy, reliability, energy efficiency, update rate, and reduced latency in position information delivery.
In this paper, we propose the exploitation of backscattering from retro-directive antenna arrays (RAAs) to address these imperatives. 
We introduce and discuss two RAA-based architectures designed for various applications, including network localization and navigation. These architectures enable swift and simple angle-of-arrival estimation by using signals backscattered from RAAs. They also leverage multiple antennas to capitalize on multiple-input-multiple-output (MIMO) gains, thereby addressing the challenges posed by the inherent path loss in backscatter communication, especially when operating at high frequencies.
Consequently, angle-based localization becomes achievable with remarkably low latency, ideal for mobile and vehicular applications. 
This paper introduces ad-hoc signalling and processing schemes for this purpose, and their performance is analytically investigated. Numerical results underscore the potential of these schemes, offering precise and ultra-low-latency localization with low complexity and ultra-low energy consumption devices.
\end{abstract}


\section{Introduction}
Towards the realization of 6G,  the integration of advanced localization and sensing capabilities holds paramount importance, enabling unprecedented levels of spatial awareness and context-driven intelligence to revolutionize communication, connectivity, and interaction in diverse domains \cite{Del:J21,SarSaeAlNAlo:20}.

For instance, in vehicular networks, the integration of localization and sensing technologies is crucial for enhancing safety, efficiency, and connectivity by enabling real-time awareness of vehicle positions, environmental conditions, and surrounding traffic dynamics \cite{WymEtAl:J17,BarEtAl:J21}. This becomes even more important when it comes to autonomous driving systems \cite{Whi:M22}: as the level of automation increases, so does the demand for enhanced accuracy, reliability, update rate, and reduced latency in position information delivery. In particular, position information must be updated several times per second to properly feed the control systems (i.e., \textit{high update rate}), and it must be extremely up-to-date to enable fast reactions of the vehicles even in high-speed contexts (i.e., \textit{extremely low latency}) \cite{SaiYilMicPerKeaSch:21,ko2021v2x,DecGueGioGuiMas:J23}. 
Similar requirements also arise across diverse types of vehicular networks, including \acp{UAV}  \cite{GueDarDju:J20a} and those involving mobile robots in industrial plants for optimizing operational efficiency and enhancing safety \cite{EhrPetSarTesMelSchAnwFraFetMar:17}.

When the localization demand comes together with stringent energy efficiency and low complexity requirements, radio backscattering represents a viable solution \cite{MotBufNep:J21,DecGuiDar:J16,ZhaWanTanEtAl:J20,DecDelMasDarCos:J18}. Backscattering enables devices, namely tags, to transmit data through the reflection of interrogation signals \cite{XuYanZha:18}. This approach offers advantages such as ultra-low power consumption and low-complexity implementations since neither active transmitters nor complete \ac{RF} chains are usually embedded in tags. As examples in vehicular contexts, backscatter radios could be used on board swarms of drones to be localized (\textit{network localization} application), or as energy-autonomous reference tags (for instance, exploiting energy harvesting technologies) deployed in the environment for the self-localization of autonomous vehicles (\textit{navigation} application). 
Regrettably, the primary drawback of backscatter radio is the significant path loss, as the signal traverses the propagation channel twice. This results in a limited operating range, often just a few meters, as observed in \ac{RFID} applications \cite{MieEtAll:11}. Such constraints are exacerbated with higher frequencies (mmWave/THz) \cite{ElaAbbAbuZheSolKai:18}, rendering traditional backscatter schemes impractical for achieving high-accuracy localization in the applications mentioned above. 

A solution to counteract the path loss is the adoption of \ac{MIMO} techniques, involving the integration of multiple antennas on both nodes engaged in the backscattering process (i.e., the emitter of the interrogation signal and the backscatter radio). However, this would require maintaining the respective beams aligned, which contradicts the low-complexity nature of backscatter radio, in which no \ac{CSI} estimation can be performed. 
In addition, in dynamic scenarios such as those involving terrestrial vehicles or \acp{UAV}, the challenges of beam alignment with moving and/or fluctuating objects are further exacerbated \cite{HeFakAle:24}, especially when considering mmWave/THz due to the pencil-like beams that can be realized at these frequencies. Therefore, to envisage the adoption of backscatter-based devices equipped with multiple antennas at high frequencies, it is crucial to develop innovative beamforming strategies capable of dynamically adapting to the movements of communicating devices, without necessitating processing capability and, ideally, with reduced complexity compared to traditional beamforming approaches.

In this regard, \acp{RAA} \cite{PasCodPusDolPal:18}, \cite{YenChu:04} deserve special consideration. These arrays possess a unique property called \emph{retrodirectivity}, enabling them to act as intelligent mirrors that reflect incoming signals back to the source's direction, without requiring explicit knowledge of the source's location \cite{MiyIto:02}. This allows for efficient exploitation of the \ac{MIMO} gain, even in backscatter radio solutions. It is worth noticing that retrodirectivity (also known as \textit{retroreflection}) can be obtained using the recently introduced \ac{RIS} technology. However, it requires the knowledge of the source's position to properly configure the \ac{RIS} phase-profile  \cite{TagMizOliSpaMas:23}.

Actually, \acp{RAA} have undergone research attention across diverse applications, encompassing satellite communications \cite{BucFusVan:16}, \cite{ZhuHuLinLi:19}, and terrestrial communication systems \cite{KarFus:98}.  However, only a few works considered \acp{RAA} for localization purposes, as for example \cite{SolPraBalAndRabKumRow:21}. In this work, the authors present a tag designed to achieve long-range capabilities by strategically reflecting incident waves back toward the reader. They leverage the large bandwidth offered by mmWave frequencies to enhance accuracy and propose the utilization of the combination of \ac{FFT} based techniques and MUSIC algorithms for estimating angles and distances. 
The detection capability of a backscattering \ac{RAA}-based \ac{RFID} system is instead investigated in \cite{HesTen:17}. Here, the authors introduce an energy-autonomous, long-range-compatible \ac{RFID} system operating at mmWave frequencies. Additionally, they present the results of experiments conducted to evaluate the device's performance and measure its detection range, showing an ultra-long range of 80 meters. In \cite{GupBro:07} a radar system exploiting an \ac{RAA} is shown. Initially emitting omnidirectional pulses, it gradually gains directivity toward the target, improving signal quality over successive pulses. 
In \cite{ZhuHuQinLiLiZenLinLig:20}, an \ac{RAA} is introduced that receives a $40\,$GHz navigation signal and accurately re-transmits a $120\,$GHz beam in the direction of the incoming wave, utilizing internal local oscillators. The authors present simulation results indicating that this antenna can track the incoming wave with a low relative error. 
 
The previous studies have been mainly focused on enabling technologies and implementation-related aspects of RAAs. None of them have systematically investigated their potential for localization networks, particularly concerning the diverse requirements of various applications. 
 
Moreover, ad-hoc and low complexity processing schemes, capable of optimum communication and localization performance in \ac{MIMO} architectures involving backscattering RAA, especially in the mmWave/THz scenarios are, to the best of the authors' knowledge, still lacking. 

In this study, we investigate the utilization of \acp{RAA} for localization in \ac{MIMO} networks comprising mobile agents, such as terrestrial vehicles, \acp{UAV}, and mobile robots in industrial environments. Specifically, we illustrate two architectures tailored to enable angle-based localization by leveraging \acp{RAA} and narrowband signals, the first suitable for network localization, where  \ac{RAA}-based devices are the mobile nodes to be localized, and the second where fixed \ac{RAA}-based devices are deployed in the environment and exploited by mobile nodes for navigation applications. To this end, we introduce an iterative technique to automatically perform beam alignment between a multi-antenna transmitting node and an \ac{RAA}, thereby enabling rapid and simple \ac{AoA} estimation, with optimum performance in terms of \ac{MIMO} gain and thus communication/estimation quality. Furthermore, fusing multiple \ac{AoA} estimates, obtained in parallel using \acp{RAA}, enables localization with significantly lower latency and higher update rate compared to currently available \acp{RTLS} with the additional advantages of exploiting narrowband signals and ultra-low energy backscattering devices. The main contributions can be summarized as follows.

\begin{itemize}
    \item We provide an extended discussion concerning the possibility of narrowband angle-based localization by leveraging \acp{RAA} in \ac{MIMO} mobile wireless networks.
    \item We propose a blind iterative scheme capable of estimating the optimal beamforming and recovering the \ac{AoA} at the transmitting/receiving antenna array with extremely low latency and avoiding complex and explicit \ac{CSI} estimation.   
    \item We characterize analytically and numerically the performance of the proposed scheme, and we introduce an ad-hoc initialization strategy to further reduce the latency in the presence of nodes' mobility.
\end{itemize}

The rest of the paper is organized as follows. In Section \ref{sec:LocRAA}, we provide the fundamental concepts of \acp{RAA}, introduce the envisioned network architectures, and discuss their key components. Moving forward, in Section \ref{sec:AoARAA}, we demonstrate how \acp{RAA} can be exploited to enable \ac{AoA} estimation, and we present a scheme capable of both deriving \acp{AoA} and extracting data bits from the backscattered signal, also addressing the case of multiple users. In Section \ref{Sec:Performance_Evaluation}, we analytically investigate the convergence of the proposed scheme and discuss the benefits of introducing tracking capabilities at the channel level. Finally, in Section \ref{Sec:Numerical Results},  we present the numerical results, highlighting the performance of our solution.

\bigskip
{\textit{\textbf{Notations and Definitions:}}
Boldface lower-case letters are vectors (e.g., $\boldx$), whereas boldface capital letters are matrices (e.g., $\boldH$). 
$\|\boldx \|$ represents the Euclidean norm of vector $\boldx$ and $\boldx^*$ is its conjugate. $\boldH^*$, $\boldH^{\top}$ and $\boldH^{\dag}$ indicate, respectively, the conjugate, the transpose and the conjugate transpose operators applied to matrix $\boldH$.  
The notation $x \sim {\mathcal{CN}}\left (m, \sigma^2 \right )$ indicates a complex circular symmetric Gaussian \ac{RV} with mean $m$ and variance $\sigma^2$, whereas $\boldx \sim {\mathcal{CN}}\left (\mathbf{m}, \boldC \right )$ denotes a complex Gaussian random vector with mean $\mathbf{m}$ and covariance matrix $\boldC$. }

\section{Localization using Retro-Directive Antenna Arrays}
\label{sec:LocRAA}
In this section, we first introduce \acp{RAA}, highlighting the key features that can be leveraged to achieve \ac{AoA} estimation and communication. Then, we discuss the possible architectures according to which we envision their utilization for localization and navigation purposes. Finally, we conclude the section by discussing the fundamental building blocks needed for the realization of the proposed system.

\subsection{Retro-directive Antenna Arrays}
An \ac{RAA} is a particular type of antenna array that operates based on the backscattering principle. The distinctive characteristic of an \ac{RAA} is that its backscattering direction corresponds to the direction of the impinging signal, also denoted as interrogation signal, which is thus retro-directed towards the transmitter (retrodirectivity property). This behaviour can be obtained, for instance, by conjugating the phase of the received signal at each antenna element of the \ac{RAA} \cite{MiyIto:02}, or using passive structures such as Van Atta arrays \cite{Van_Atta_59}, self-conjugating metasurfaces \cite{KalSee20}. A simpler alternative approach is to compose the array using $M$ static reflectors each differently configured to retro-reflect the impinging signal from a specific direction  \cite{TagMizOliSpaMas:23}. However, compared to \acp{RAA}, this solution offers a reduced gain of $M$ and a limited set of $M$ working directions.

Remarkably, backscattering towards the transmitter is achieved without the need to estimate the \ac{AoA} of the signal impinging on the \ac{RAA}, as its behaviour is not based on phase shifters as in conventional antenna arrays. This approach ensures minimal complexity, low energy consumption, and eliminates the need for control signals between the transmitter and the \ac{RAA}-equipped device. Additionally, leveraging the architecture proposed in \cite{DarLotDecPas:J23}, data can be embedded in the backscattered signal (e.g., containing the device \ac{ID}), thus establishing a communication link between the \ac{RAA}-equipped device and a receiver possibly integrated into the same device that houses the transmitter.
Therefore, an array-equipped receiver, co-located with the transmitter generating the interrogating signal, can collect the response signal backscattered by the \ac{RAA} and exploit it to perform data demodulation (e.g., to identify the device) and \ac{AoA} estimation, which can be further exploited for localization purposes, as investigated in this paper.

Concluding this short introduction on \acp{RAA}, it is important to note that the localization of \acp{UE} is commonly achieved through the deployment of a set of reference nodes, referred to as \textit{anchors}, positioned at fixed known locations. These nodes, along with the \acp{UE} themselves, interact to estimate the characteristics of position-dependent signals, such as the \ac{AoA} \cite{DarCloDju:J15}. In the following, we will illustrate how \acp{RAA} can be exploited both at the \ac{UE} and anchor sides, leading to different architectural solutions, each with their own set of advantages and disadvantages.

\begin{figure}[t]
\centerline{\includegraphics[width=0.8\columnwidth]{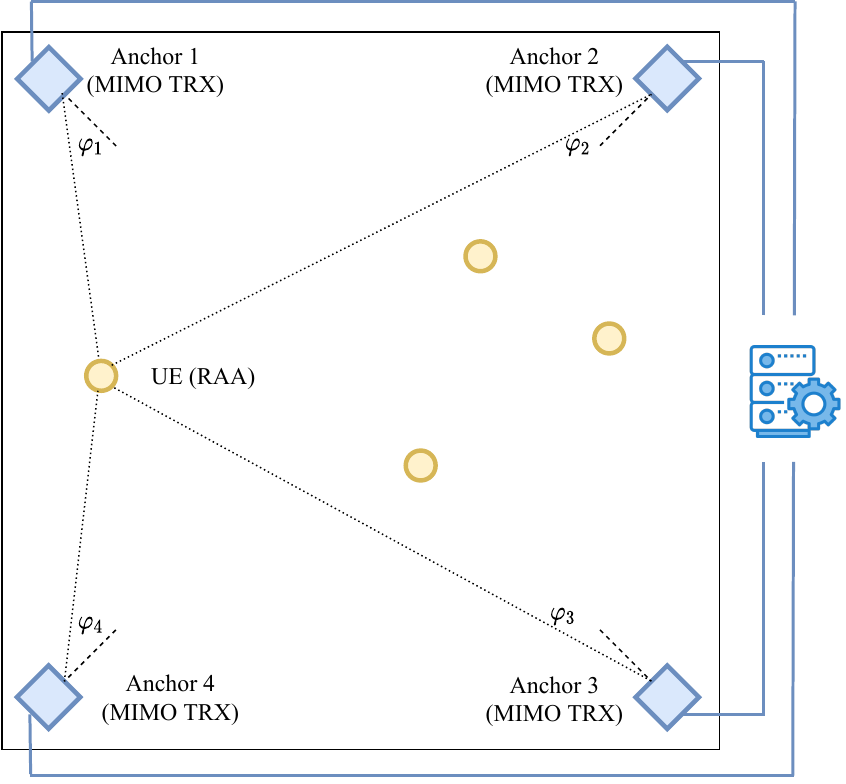}}
	\caption{Localization of RAAs-equipped mobile UEs (yellow circles) by using array-equipped anchors (blue squares). Anchors are also labelled a MIMO transceivers (TRX).}
	\label{fig:SCM-LOCue}
\end{figure}

\subsection{Architectures}\label{sec:schemes}
\begin{figure}[t]
\centerline{\includegraphics[width=0.73\columnwidth]{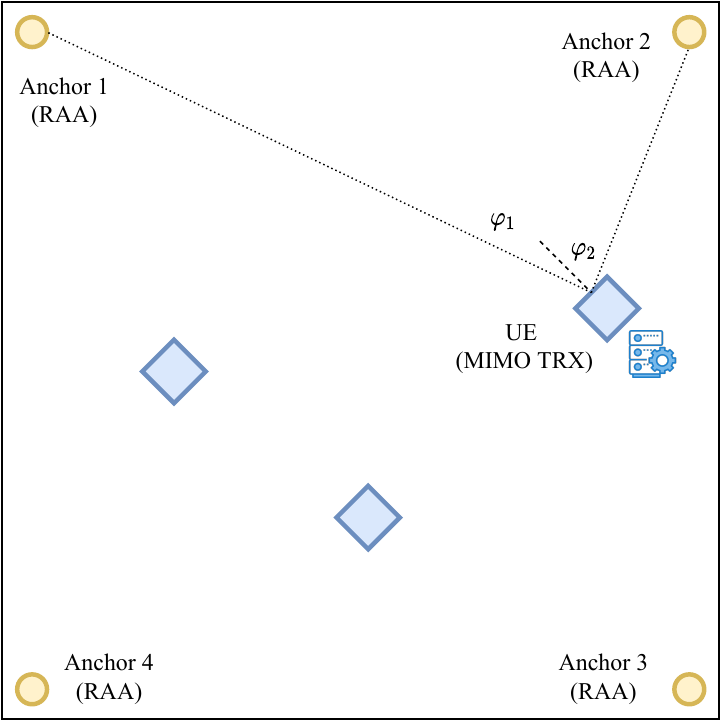}}
	\caption{Navigation of array-equipped mobile UEs (blue squares) through the interaction with RAAs (yellow circles) used as anchor nodes.  \acp{UE} are also labelled MIMO transceivers (TRX).}
	\label{fig:SCM-LOCanc}
\end{figure}

We designate \textit{Architecture 1} as the configuration where \acp{RAA} are employed on mobile \acp{UE}, while anchor nodes use conventional antenna arrays (see Fig.~\ref{fig:SCM-LOCue}). Conversely, \textit{Architecture 2} involves \acp{RAA} for anchors while \acp{UE} are equipped with conventional antenna arrays (see Fig.~\ref{fig:SCM-LOCanc}). In the following, we introduce the fundamental building blocks necessary for realizing communication and \ac{AoA} estimation by leveraging the availability of \acp{RAA}-equipped devices, either anchors or \acp{UE}, in the scenario.

\subsubsection{Architecture 1 - \ac{RAA} at the \ac{UE} Side}
Each mobile \ac{UE} is equipped with an \ac{RAA} (Fig.~\ref{fig:SCM-LOCue}), while each anchor is equipped with a conventional array and is configured to transmit an interrogation signal. This signal is backscattered by the \acp{RAA}, which also embed their \acp{ID}, and subsequently received by the same anchor. The anchor can then estimate the \acp{AoA} of the signals received by the \acp{RAA}, and extract their \ac{ID} using the scheme introduced in the following sections. Such information is then passed to a localization engine, which is in charge of estimating the \acp{UE}' positions based on the data received by all anchors.  Once a sufficient number of \ac{AoA} measurements are collected for a given \ac{RAA}, the localization engine can localize the \acp{UE}, knowing the anchors' positions and orientations. For example, at least two {\acp{AoA}} per {\ac{UE}} are needed for unambiguous 2D localization.
Clearly, the larger the number of \ac{AoA} measurements for each \ac{UE}, the better will be the localization accuracy. 

To deal with measurement errors, the localization engine can fuse different \acp{AoA} estimates by leveraging, for instance, standard tools such as least squares or particle filtering.\footnote{The reader can refer to \cite{DarCloDju:J15}.}
Different anchors can access simultaneously the \acp{UE} by exploiting, for example, different subcarriers, according to an \ac{OFDM} signalling, assuming a flat frequency response of the \ac{RAA} within the signal bandwidth. 
The discrimination of the backscatter components associated with different \acp{UE}, can be realized by exploiting the antenna array spatial selectivity, as will be explained in Sec. \ref{Sec:Extension_to_Multiple_Users}.

This architecture is primarily tailored for network localization applications, enabling the network to gain comprehensive insights into the positions of all \acp{UE} within the scenario of interest. This encompasses a wide set of use cases such as asset and personnel tracking, logistics management, and more.
The primary advantage of this architecture lies in the potential for remarkably simple \acp{UE}, as they require no processing capabilities on their side. Essentially, the \ac{UE} solely engages in backscattering the incoming signal using the \ac{RAA} and incorporating its \ac{ID} in the reflected signal. Consequently, there is no need for complex \ac{RF} chains or baseband components within the \acp{UE}, resulting in reduced costs and complexity as well as the possible exploitation of energy harvesting techniques making the \acp{UE} energy autonomous.

\subsubsection{Architecture 2 - \ac{RAA} at the Anchor Side}
This architecture entails equipping each anchor node with an \ac{RAA}, while \acp{UE} utilize a conventional antenna array (Fig.~\ref{fig:SCM-LOCanc}). In contrast to Architecture 1, in this scenario, it is the anchor that responds to the interrogation signal emitted by the \ac{UE}, leveraging retrodirectivity. Consequently, each \ac{UE} must detect the presence of the \ac{RAA}-equipped anchors, extracting their \acp{ID}, and estimate the \acp{AoA} of the retro-directed signals relative to its local coordinate system. A decentralized localization engine operates within each \ac{UE}, enabling the estimation of its position.

This architecture is primarily suited for navigation purposes, resembling {GNSS}-like positioning, as the position computation occurs at the \ac{UE} level. When considering vehicular networks, this architecture is particularly suited for the self-localization of autonomous vehicles (where no complexity constraints usually arise on-board vehicles) using simple, low-cost and possibly energy autonomous reference tags (equipped with \acp{RAA}) deployed in the environment. In general, this architecture allows for reducing the complexity and cost of anchor infrastructure, which often serve as the primary barriers to introducing high-accuracy positioning systems like those based on \ac{UWB} technology \cite{DarCloDju:J15}.

The different \acp{UE} can access simultaneously the \ac{RAA}-based anchors by exploiting, for example, different subcarriers of an \ac{OFDM}-based system. As it will be clearer in Sec. \ref{Sec:Extension_to_Multiple_Users}, a given \ac{UE} can address simultaneously different anchors thanks to the spatial discrimination allowed by the use of antenna arrays with a large number of antennas and the proposed scheme. 

A key distinction from Architecture 1 is that, in this scenario, localization information is immediately available at the \ac{UE} itself. This minimizes latency in scenarios where \acp{UE} require prompt awareness of their positions. This is crucial when utilizing position information to feed navigation engines, especially in autonomous driving applications involving vehicles, \acp{UAV} and mobile robots.
This architecture can be further improved by enabling the \ac{RAA}-based anchors to transmit not only their \ac{ID} but also their coordinates. This additional information can be utilized by the \ac{UE} for localization without requiring prior knowledge of the anchors' deployment layout. Furthermore, in a practical system, multiple anchors can be seamlessly added without the need to update the anchor database at the \ac{UE} side.
It is worth noticing that, differently from Architecture 1,  the orientation of the \ac{UE} could be unknown and hence it must be estimated to obtain the \ac{UE}'s absolute coordinates. To this purpose, also additional sensors such as an inertial and/or a compass can be considered.

\subsection{Building Blocks}

To effectively implement the proposed architectures, specific building blocks and dedicated methods are required. 
It is important to note that in both architectures, devices equipped with standard antenna arrays serve as both transmitters, sending the interrogation signal, and receivers, collecting the backscattered signal. Indeed, the adoption of multiple antennas is required to counteract the unfavourable path loss typical of backscatter communication as well as to allow the estimation of the signal's \ac{AoA}. Thus, the same antenna array can be used for transmission and reception by exploiting a full-duplex radio implementation, as commonly considered in monostatic radar applications \cite{BarLiyHeiRii:J21}. In the following discussion, we will refer to  \textit{\ac{MIMO TRX}} and \textit{\ac{RAA}-based device} (or simply \ac{RAA}) as follows: in Architecture 1, anchors and \acp{UE}, and in Architecture 2, \acp{UE} and anchors, respectively.

The primary challenge in both architectures is determining the beamforming vector at the \ac{MIMO TRX} to direct the interrogation signal towards the \ac{RAA}, whose direction remains unknown prior to \ac{AoA} estimation. To tackle this challenge, we propose an iterative procedure inspired by \cite{DarLotDecPas:J23}, leveraging the distinctive capability of \acp{RAA} to reflect the signal in the same direction it was received. This approach allows for the automatic determination of the optimal beamforming vector at the transmitter's side without requiring explicit channel estimation and signaling. Remarkably, the determination of the beamforming vector intrinsically provides the \ac{AoA} of the backscattered signal, subsequently used for localization. Clearly, since a \ac{MIMO} link is finally established, the communication benefits from the \ac{MIMO} gain to mitigate the high path loss. 
This procedure will be detailed in Sec.~\ref{sec:AoARAA}. 

\section{AoA Estimation Based on \acp{RAA}}\label{sec:AoARAA}

In this section, we outline a procedure for the \ac{MIMO TRX} to estimate the \ac{AoA} of the signal backscattered by an \ac{RAA} using a blind iterative approach, guiding the \ac{MIMO TRX} to direct its interrogation signal towards the \ac{RAA}. Subsequently, the scheme is extended to accommodate multiple \ac{RAA}-based devices, allowing for simultaneous estimations.

\subsection{System Model}\label{sec:SysMod}
\begin{figure}[t]
\centerline{\includegraphics[width=\columnwidth]{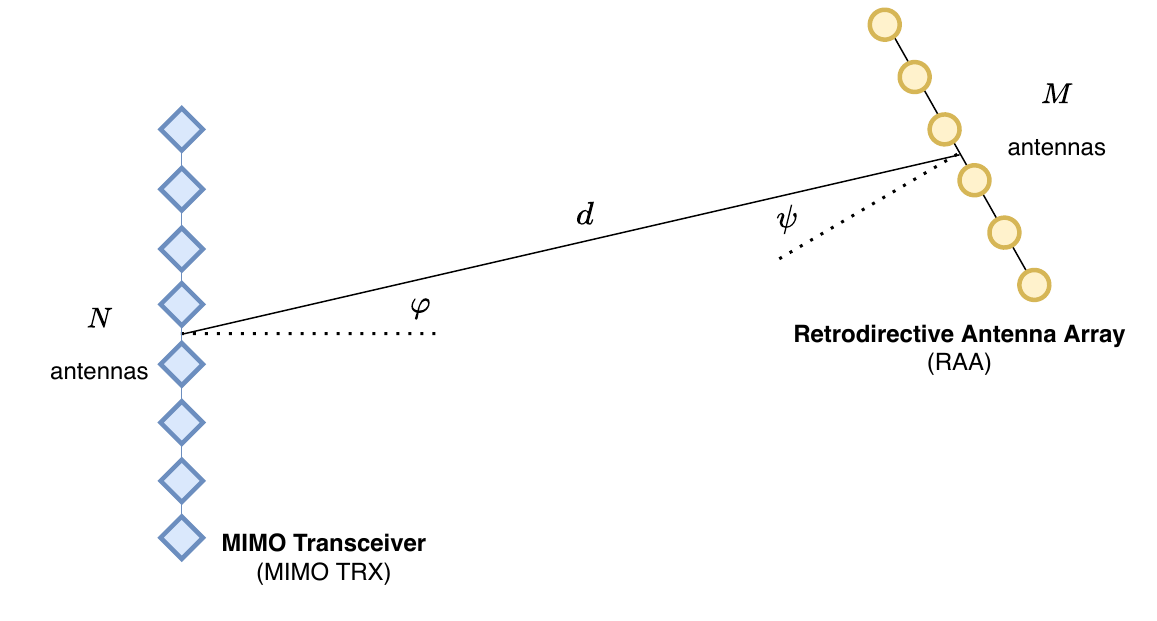}}
	\caption{Geometry of the scenario. MIMO TRX equipped with a $N$-antennas ULA; RAA composed of $M$ antennas organized as ULA.}
	\label{fig:SCMloc}
\end{figure}

We start by considering only two nodes, namely, a \ac{MIMO TRX} and an \ac{RAA}. The \ac{MIMO TRX} is equipped with a \ac{ULA} comprising $N$ elements, capable of full-duplex communications. The \ac{RAA}, on the other hand, is realized as a uniform linear array consisting of $M$ elements, without processing capabilities (refer to Fig.~\ref{fig:SCMloc}). Here, we focus on uniform linear arrays for simplicity of explanation, although various array layouts can be explored, as demonstrated in the numerical results. Both arrays are located in the far-field region of each other, separated by a distance $d$. The \ac{AoD} of the signal emitted by the \ac{MIMO TRX}, when directed towards the \ac{RAA}, is denoted as $\varphi$, while the \ac{AoA} of the signal received at the \ac{RAA} from the \ac{MIMO TRX} is denoted as $\psi$.     

The scheme we propose involves transmitting from the \ac{MIMO TRX} to the \ac{RAA} using a certain beamforming vector $\boldx\in \mathbb{C}^{N \times 1}$, ideally aligned with direction $\varphi$. 
Thanks to the retrodirectivity capability, the signal impinging the \ac{RAA} is backscattered towards the direction of arrival (i.e., angle $\psi$). The backscattered signal is then received by the \ac{MIMO TRX} by exploiting a full-duplex radio. This signal exchange, from the \ac{MIMO TRX} to the \ac{RAA} and back, takes place iteratively, once every $T$ seconds corresponding to the symbol time. The time axis is thus segmented into discrete intervals, indexed by $k$.

At the startup, the optimum beamforming vector for the link with the \ac{RAA} is not known by the \ac{MIMO TRX}, which therefore randomly generates a unit norm beamforming vector $\boldx[0]$.
As will be detailed in Sec.~\ref{sec:algorithm},  at the end of each time interval $k$, with $k\ge 1$, the beamforming vector $\boldx[k]$ will be iteratively updated with a scheme allowing convergence towards the direction $\varphi$, finally establishing an optimal \ac{MIMO} link between the two nodes.

\begin{figure}[t]
	\centerline{\includegraphics[width=0.7\columnwidth]{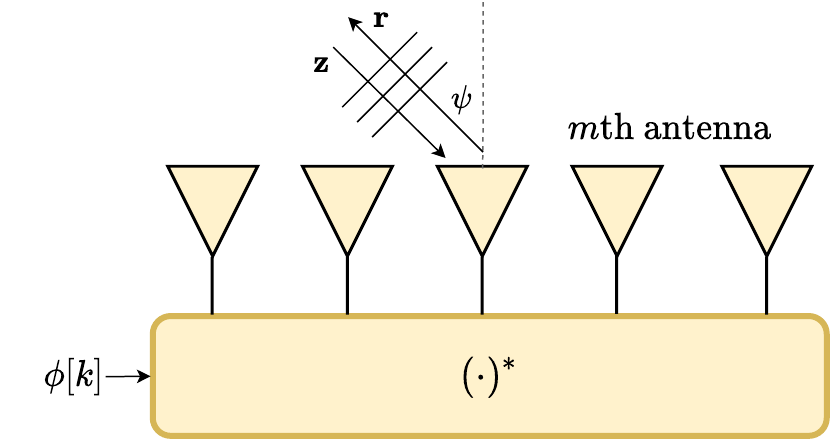}}
	\caption{Schematic representation of an RAA organized as ULA.}
	\label{fig:SCM}
\end{figure}

Let $\sqrt{\Ptx}\,\boldx[k]\in \mathbb{C}^{N \times 1}$ be the vector containing the signal transmitted by the $N$ elements of the \ac{MIMO TRX}'s antenna array, where $\Ptx$ is the transmitted power and $\boldx[k]$ is the unit norm beamforming vector at the generic time interval $k$.
At the other end of the communication link, consider a plane wave impinging on the \ac{RAA}, the schematic representation of which is shown in Fig.~\ref{fig:SCM}, with an angle $\psi$ with respect to its normal direction.
At the $m$-th RAA antenna, the impinging wave accumulates a phase shift $\theta_m$, with respect to the first antenna, given by 
\begin{equation}
	\theta_m=\frac{2\pi}{\lambda}m\Delta\sin\psi
\end{equation}
for $m=0,1, \ldots, M-1$, where $\Delta$ is the inter-antenna spacing and $\lambda$ is the wavelength. 
By introducing the noise generated by the \ac{RAA}, which is present in case it is implemented using active components  \cite{BouMagMes:12,ZhaDaiCheLiuYanSchPoo:22}, the discrete-time signals at the input of the $M$  antennas in the $k$-th time interval can be expressed by the vector 
\begin{equation}\label{eq:received_Vector}
	\boldz[k]=\alpha[k]  \left[1,\, e^{\jmath \theta_1},\,  \ldots,\,  e^{\jmath \theta_{M-1}} \right]^{\top} + \boldeta[k]
\end{equation}
where  $\alpha[k]$ is the signal at the first antenna, ${\boldeta[k] \in \mathbb{C}^{M \times 1}}$ is the \ac{AWGN}, with ${\boldeta[k] \sim {\mathcal{CN}}\left (\mathbf{0}, \sigma_{\eta}^2 \boldI_M \right )}$. Note that ${\sigma_{\eta}^2=\kappa T_0 \Fscm\, W}$, where $\kappa$ represents the Boltzmann constant, $T_0=290\,$K denotes the reference temperature, $\Fscm$ stands for the RAA's noise figure, and $W$ indicates the signal bandwidth. Regarding the latter, we consider a narrowband transmission, such as a resource block in an \ac{OFDM} communication scheme. 

To realize the retrodirectivity property, the phase profile of the reflected signal along the array must be opposite of that of the impinging signal. Therefore, the vector $\boldr[k]$ representing the signal backscattered by the \ac{RAA} during the same time interval should be \cite{MiyIto:02}
\begin{equation}
\label{eq:SCMmodel}
\boldr[k]=g \, \boldz^*[k] 
\end{equation}
where $g$ is the gain of the \ac{RAA} ($g<1$ if passive, meaning that no amplifier is used). 
From an implementation viewpoint, the conjugation in \eqref{eq:SCMmodel} can be obtained explicitly by means of an active circuit based on the superheterodyne principle \cite{MiyIto:02,AllLeoIto03}. Alternatively, passive solutions can be adopted based on Van Atta arrays \cite{ShaDia:60} or ad-hoc designed metasurfaces, which yield an equivalent result in the far-field region of the \ac{RAA} \cite{KalSee20}. 

In addition to conjugating the received signal, we assume that the \ac{RAA} introduces a phase shift $\phi[k]$ into the backscattered signal during the $k$-th time interval. This phase shift conveys information from the \ac{RAA} to the \ac{MIMO TRX} during that interval. For instance, this information could include a unique signature of the specific node, such as its \ac{ID}. Note that the phase shift $\phi[k]$ is the same across all antennas, thereby preserving the retrodirectivity property.  
The vector representing the  signal reflected by the \ac{RAA} becomes therefore
\begin{equation}
\label{eq:SCMmodel_with_info}
\boldr[k]=g \, e^{\jmath \phi[k]}  \, \boldz^*[k]  \, .
\end{equation}

By denoting the channel matrix between the \ac{MIMO TRX} and the \ac{RAA} with $\boldH \in \mathbb{C}^{M \times N}$, the signal received by the \ac{RAA} at time instant $k$ is given by
\begin{equation}
\boldz[k]=\sqrt{\Ptx}\,  \boldH \, \boldx[k-1] + \boldeta[k]
\end{equation}
and the retro-directed signal, according to  \eqref{eq:SCMmodel_with_info}, is
\begin{align}
\boldr[k]=&g \, e^{\jmath \phi[k]} \, \boldz^*[k] \nonumber \\
=& \sqrt{\Ptx}\, g\,  e^{\jmath \phi[k]} \, \boldH^* \, \boldx^*[k-1] + g\, \boldeta^*[k]\, .
\end{align}
Considering a free-space \ac{LOS} scenario, and denoting with $\Gap$ and $\Gscm$ the gain of each antenna element at the \ac{MIMO TRX} and \ac{RAA}, respectively, the channel matrix $\boldH$ takes the form
\begin{align}
\label{eq:H_ang}
\boldH(\varphi,\psi)&=\sqrt{\Gap \, \Gscm}\frac{\lambda}{4\pi d} \underbrace{
\begin{bmatrix}
1 \\ 
e^{-\jmath \frac{2 \pi}{\lambda} \Delta \sin\psi} \\
\vdots \\
e^{-\jmath \frac{2 \pi}{\lambda} (M-1)\Delta \sin\psi}
\end{bmatrix}}_{\tilde{\boldu}(\psi)\in \mathbb{C}^{M \times 1}}\nonumber \\
&\quad\quad\times
\underbrace{
\begin{bmatrix}
1 & 
e^{-\jmath \frac{2 \pi}{\lambda} \Delta \sin\varphi} &
\ldots &
e^{-\jmath \frac{2 \pi}{\lambda} (N-1)\Delta \sin\varphi}
\end{bmatrix}}_{\tilde{\boldv}^{\top}(\varphi)\in \mathbb{C}^{1 \times N}}
 \nonumber \\
&=\sqrt{\Gap \, \Gscm}\frac{\lambda}{4\pi d} \tilde{\boldu}(\psi)\tilde{\boldv}^{\top}(\varphi) 
\end{align}
where we have highlighted with $\boldH(\varphi,\psi)$ the dependence of the channel on the \ac{AoD} $\varphi$ at the \ac{MIMO TRX} and \ac{AoA} $\psi$ at the \ac{RAA}. It is worth noting that the channel matrix $\boldH(\varphi,\psi)$ depends on the angles $\varphi$ and $\psi$ (i.e., on the geometry of the scenario), regardless of whether the beamforming vector at the \ac{MIMO TRX}'s side $\boldx[k]$ corresponds to the beam steering vector in the direction $\varphi$ (i.e., the optimal direction to convey power towards the \ac{RAA}) or not.

By defining the vectors $\boldu(\psi)=\tilde{\boldu}(\psi)/{\sqrt{M}} \in \mathbb{C}^{M \times 1}$ and $\boldv(\varphi)=\tilde{\boldv}^*(\varphi)/{\sqrt{N}} \in \mathbb{C}^{N \times 1}$, it results 
\begin{align}\label{eq:LOSchannel}
\boldH(\varphi,\psi)=\sqrt{NM\Gap \, \Gscm}\frac{\lambda}{4\pi d} \boldu(\psi)\boldv^\dag(\varphi) \, 
\end{align}
which has rank one since obtained as an outer product of two vectors $\boldu$ and $\boldv^*$.  Notice that $\boldu$ and $\boldv$ are, respectively, the top left and right eigenvectors of matrix $\boldH(\varphi,\psi)$ and hence give the optimum beamforming vectors at the \ac{RAA} side and at the \ac{MIMO TRX}.\footnote{Since $\boldH(\varphi,\psi)$ has rank one, $\boldu$ and $\boldv$ correspond to the only left and right eigenvectors associated with non-zero eigenvalues. In this sense, they are referred to as the top eigenvectors.} 
We can introduce the \ac{SVD} of $\boldH(\varphi,\psi)$ as $\boldH(\varphi,\psi)=\boldU \boldSigma \boldV^\dag$
where $\boldSigma$ has a singular non-zero entry 
\begin{equation}\label{eq:lambda1}
\sigma_1=\frac{\sqrt{NM\Gap \, \Gscm}\lambda}{4\pi d}
\end{equation}
in its first element, $\boldU$ has $\boldu(\psi)$ as the first eigenvector (i.e., first column), and $\boldV$ has $\boldv(\varphi)$ as the first eigenvector (i.e., first column).

Assuming channel reciprocity, at the \ac{MIMO TRX} side the received signal at time interval $k$, consisting of the feedback of the signal transmitted in the last time interval, is given by\footnote{For simplicity, in this model we do not consider clutter, that is, the signal backscattered by the environment and not modulated by the \ac{RAA}. The reader can refer to \cite{DarLotDecPas:J23} for a discussion concerning the modelling of the clutter and its impact on communication.}
\begin{align} \label{eq:byk}
\boldy[k]=& \sqrt{\Ptx} \, g \, e^{\jmath \phi[k]} \, \boldH^{\top} (\varphi,\psi) \, \boldH^{*}(\varphi,\psi) \,\boldx^*[k-1] \nonumber \\
& + g\, \boldH^{\top}(\varphi,\psi) \, \boldeta^*[k] + \boldw[k] 
\end{align}
with $\boldw[k] \sim {\mathcal{CN}}\left (\mathbf{0}, \sigma_w^2 \boldI_N \right )$ being the \ac{AWGN} at the receiver, and $\sigma_w^2=\kappa T_0 \Fap   \, W$, where $\Fap$ represents the \ac{MIMO TRX}'s noise figure. 
By defining ${\boldA(\varphi)=\sqrt{\Ptx} \, g\,  \boldH^{\dag}(\varphi,\psi) \boldH(\varphi,\psi) \in \mathbb{C}^{N \times N}}$, according to \eqref{eq:LOSchannel} it results
\begin{equation} \label{eq:boldA_LOS}
\boldA(\varphi)=\sqrt{\Ptx} \, g N M \Gap  \Gscm \left(\frac{\lambda}{4 \pi d}\right)^2 \boldv(\varphi) \boldv^{\dag}(\varphi)
\end{equation}  
which depends only on the angle $\varphi$, thanks to the retrodirectivity property of the \ac{RAA}. 
Moreover, $\boldA$ is proportional to the (modified) round-trip channel $\boldH^{\dag}(\varphi,\psi) \boldH(\varphi,\psi)$, rather than the (true) round-trip channel $\boldH^{\top}(\varphi,\psi) \boldH(\varphi,\psi)$ as in conventional backscatter communications. Consequently, the eigenvectors of $\boldA$  are identical to the right-eigenvectors of $\boldH(\varphi,\psi)$.
From \eqref{eq:byk} we have
\begin{equation} \label{eq:bykbis}
\boldy[k]=e^{\jmath \phi[k]} \, \boldA^*(\varphi)\,  \boldx^*[k-1] + \boldn^*[k]
\end{equation}  
where we have defined $\boldn^*[k]=g\, \boldH^{\top}(\varphi,\psi) \, \boldeta^*[k]+\boldw[k]$. 
Substituting \eqref{eq:boldA_LOS} into \eqref{eq:bykbis} we obtain \eqref{eq:yk_v}. 
In case of perfect beamforming, i.e.,  $\boldx[k-1]=\boldv(\varphi)$,  the useful term $\tilde{\boldy}[k]$ in \eqref{eq:yk_v} becomes
\newcounter{MYtempeqncnt1}
\begin{figure*}[ht!]
\normalsize
\setcounter{MYtempeqncnt1}{\value{equation}}
\begin{align} \label{eq:yk_v}
&\boldy[k]= \underbrace{\sqrt{\Ptx} \, g \, e^{\jmath \phi[k]}  M\! N \Gap  \Gscm\!\left(\frac{\lambda}{4 \pi d}\right)^{\!2} \!\!\boldv^*(\varphi) \boldv^{\top}\!\!(\varphi) \boldx^*[k-1]}_{\tilde{\boldy}[k]:\,\, \text{useful term}} + \underbrace{g\, \sqrt{M \!N \Gap  \Gscm} \frac{\lambda}{4 \pi d} \boldv^*(\varphi) \boldu^{\dag}(\psi)\, \boldeta^*[k] + \boldw[k] 
}_{\text{noise term}} 
\end{align}
\hrulefill
\vspace{-0.4cm}
\end{figure*}
\begin{equation}
\tilde{\boldy}[k]=\sqrt{\!\Ptx} \, g \, e^{\jmath \phi[k]}  M N \Gap  \Gscm \left(\frac{\lambda}{4 \pi d}\right)^{2} \boldv^*(\varphi)
\end{equation}
since $\boldv^{\top}\boldv^*=1$, thus corresponding to a plane wave of power proportional to $\Ptx \, M^2 N^2 / \mathcal{L}^2{(d)}$ impinging with \ac{AoA} $\varphi$, where 
\begin{equation}\label{eq:pathloss}
\mathcal{L}{(d)}=\left(\frac{4\pi d}{\lambda}\right)^2
\end{equation}
is the link loss for a distance $d$ in free space.
Notice that due to the backscattering communication type, the path loss increases with the distance to the power of four, as happens in \ac{RFID} systems \cite{DecGuiDar:J14}. On the other hand, thanks to the adoption of multiple antennas on both sides, such large path loss can be compensated by increasing the number of antenna elements $N$ and $M$ at the \ac{MIMO TRX} and \ac{RAA}, respectively (beamforming gain).

In the following section, we show how the beamforming vector $\boldx[k]$ at the \ac{MIMO TRX} side can be steered in an iterative way until converging to the channel eigenvector $\boldv(\varphi)$ pointing towards the \ac{RAA}, thus achieving the required property without resorting to explicit channel estimation. Thanks to the estimation of the optimum beamforming vector, the presence of the RAA can be detected, and the \ac{AoA} of the backscattered signal is intrinsically obtained.

\subsection{RAA Detection and AoA Estimation}\label{sec:algorithm}

The scheme we propose for \ac{RAA} detection, communication, and \ac{AoA} estimation operates iteratively.
Specifically, in the ${(k-1)}$-th time interval (i.e., during the ${(k-1)}$-th iteration of the proposed scheme), where $k\geq 1$, the following operations are performed:
\begin{enumerate}
\item The \ac{MIMO TRX} transmits $\sqrt{\Ptx} \, \boldx[k-1]$, which is the current version of the beamforming vector.  Initially, for $k=1$, the \ac{MIMO TRX} lacks knowledge of the optimal beamforming vector for the link with the \ac{RAA}. Thus, the \ac{MIMO TRX} randomly generates a unit norm beamforming vector $\boldx[0]$.
\item The \ac{RAA} receives ${\boldz[k]=\sqrt{\Ptx} \, \boldH \, \boldx[k-1] + \boldeta[k]}$ and backscatters it in the direction of arrival (thanks to the conjugation operation). Additionally, it introduces a phase modulation based on the data to be transmitted to the \ac{MIMO TRX}, resulting in the reflected signal ${\boldr[k]=g \, e^{\jmath \phi[k]} \boldz^*[k]}$. \item The \ac{MIMO TRX} receives the retro-directed and modulated response ${\boldy[k]=e^{\jmath \phi[k]} \, \boldA^*(\varphi)\,  \boldx^*[k-1] + \boldn^*[k]}$ from the \ac{RAA} at time interval $k$, as described by \eqref{eq:bykbis}.
\item At the \ac{MIMO TRX}, a normalized and conjugated version of the received vector $\boldy[k]$ is computed, which is ${\boldx[k]={\boldy^*[k]}/{\left\| \boldy[k] \right\|}}$, and used as the updated beamforming vector $\boldx[k]$ in the subsequent time interval. 
\item If the \ac{SNR} experienced by the \ac{MIMO TRX} exceeds a specific detection threshold $\eta_1$, that is, ${\gamma[k]=\|\boldy[k]\|^2/\sigma_w^2>\eta_1}$, and there is no significant increase in received power compared to the previous iteration, i.e., $\gamma[k]/\gamma[k-1]<\eta_2$, where $\eta_1$ and $\eta_2$ are suitably tuned thresholds, then it is likely that the proposed scheme has converged; thus, the \ac{RAA} is detected. We denote with $\bar{k}$ the time interval at which the detection takes place. Differently, if the conditions involving the comparison with thresholds $\eta_1$ and $\eta_2$ are not met, the process is repeated from step 1.

\item An \ac{AoA} estimate $\widehat{\varphi}$ of $\varphi$ concerning the detected \ac{RAA}  is obtained as\footnote{An array spacing $\Delta=\lambda/2$ is assumed.}
\begin{equation}\label{eq:estimation}
\widehat{\varphi}\!=\!\arcsin\!\left( \frac{2}{N}\, \operatorname{argmax}_{i \in \{1,2, \ldots , N \}}\left\{| q_i |\right\} \right) 
\end{equation}
where $\mathbf{q}=\{q_i\}$ is the \ac{DFT} of the beamforming vector corresponding to the detected RAA, that is
\begin{equation}
\mathbf{q}=\operatorname{DFT}\left[\boldy[\bar{k}]\right]\, .
\end{equation}
\item Steps 1-3 are repeated $K$ times (the length of the \ac{ID}), by using always the same beamforming vector $\boldx[\bar{k}]$, to extract the information data by correlating the current received vector with the previous beamforming vector $\boldx^{\dag}[k-1]$, thus forming the decision variable ${u[k]= \boldx^{\dag}[k-1] \, \boldy^*[k]}$.
\end{enumerate}

The decision on the modulation symbol conveyed by $\phi[k]$
at the $k$-th time interval can be obtained with a proper demodulation scheme applied to $u[k]$, depending on the modulation alphabet. In fact, in the absence of noise, it is easy to show from \eqref{eq:bykbis} that $u[k]= \boldx^{\dag}[k-1] \, \boldy^*[k] =e^{-\jmath \phi[k]}$, being $\boldx^{\dag}[k-1] \, \boldx[k-1]  =1$.
The meaningful demodulation of the data (i.e., ID) transmitted by the \ac{RAA} would require the \ac{MIMO TRX} to have a coarse synchronization with the \ac{RAA}, in order to start sending the interrogation signal at the beginning of the backscattered data packet. To address this synchronization issue, we assume that all the \acp{RAA} continuously backscatter any incoming signal, modulating it according to a specific sequence of length $K$ symbols repeated cyclically. This sequence uniquely identifies the node, thus representing its \ac{ID}. 
For instance, each \ac{RAA} can use a periodic \ac{PN} sequence of length $K$ as \ac{ID}, continuously transmitted through backscatter modulation.

In the absence of noise and data, the processing presented corresponds to the  \emph{Power Method}, or \emph{Von Mises Iteration}, which allows to estimate the strongest eigenvector of a square matrix $\boldA$, and it is described by the recursive relation \cite{GolVor20}
\begin{equation}
    \boldx[k]=\frac{\boldA \, \boldx[k-1]}{ \left\| \boldA \, \boldx[k-1]  \right\|}
\end{equation}
where $\boldx[0]$ can be either an approximation of the top (dominant) eigenvector, if available, or a random unit norm vector. It results that, for $k\rightarrow \infty$, $\boldx[k]$ converges to the top eigenvector. 
Thus, $\boldx[k]$ tends to the direction $\varphi$ of the top eigenvector of the \textit{modified round-trip channel} $\boldA(\varphi)$ (i.e., $\boldH^{\dag} \boldH$), i.e., the estimated angle converges to the \ac{AoA} of the signal from the \ac{RAA}. 

Summarizing, the iterative process here proposed allows the \ac{MIMO TRX} to transmit according to the optimum beamforming vector pointing towards the \ac{RAA}. Therefore, the \ac{AoD} of the signal from the \ac{MIMO TRX} coincides with the \ac{AoA} of the signal received from the \ac{RAA}, thus enabling angle-based localization. This is realized by exploiting the processing gain offered by the $N$ antennas at the \ac{MIMO TRX} and the $M$ antennas at the \ac{RAA}, but without explicit channel estimation, and adopting backscattering.

\subsection{Extension to Multiple RAA-based Devices}
\label{Sec:Extension_to_Multiple_Users}

\begin{figure}[t]
	\centerline{\includegraphics[width=0.99\columnwidth]{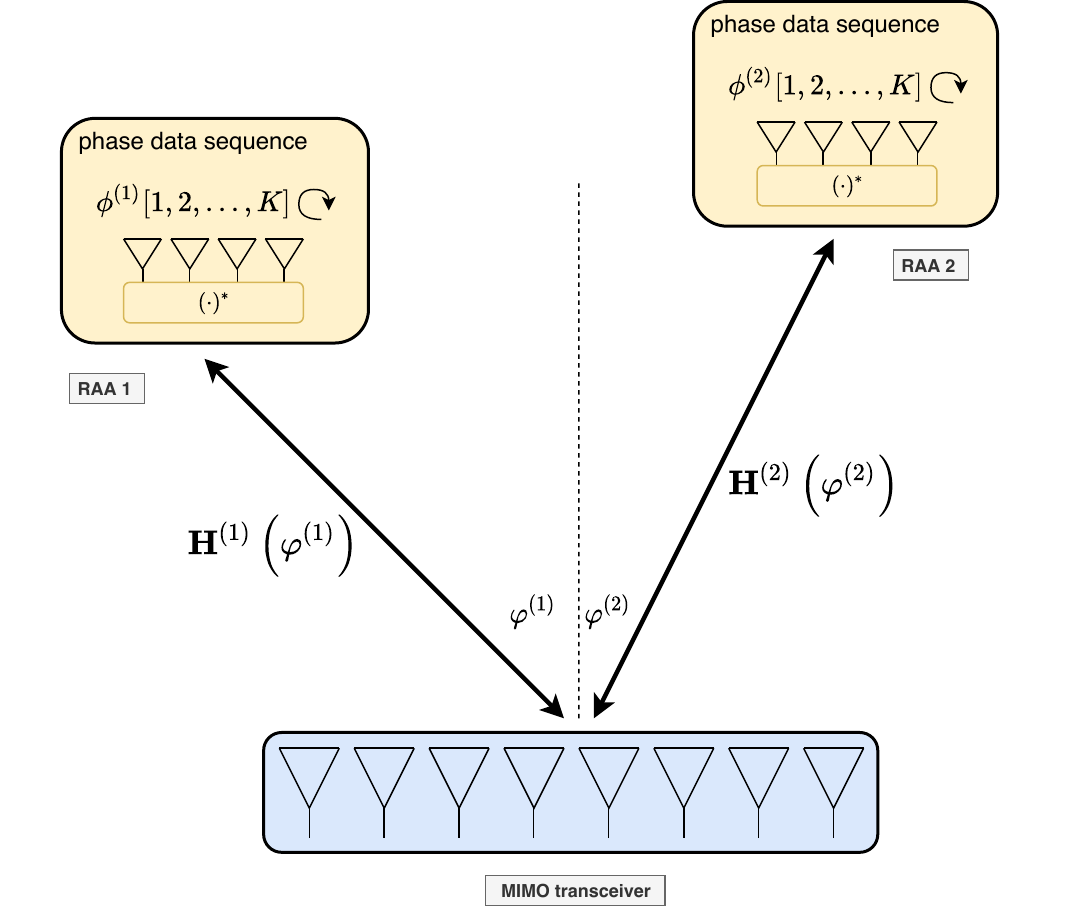}}
	\caption{Communication and AoA estimation with multiple RAAs sending packets of $K$ bytes cyclically.}
	\label{fig:SCM-multiUE}
\end{figure}

Now, let's consider $P$ \ac{RAA}-based devices present in the environment. These devices can either be mobile \acp{UE} in Architecture 1 or fixed anchors in Architecture 2. In the same environment, there are also devices equipped with a standard antenna array (\ac{MIMO TRX}), which emit interrogation signals to detect the presence of the $P$ \ac{RAA}-based devices and estimate their corresponding \acp{AoA} (see Fig.~\ref{fig:SCM-multiUE}). 

In \ac{LOS} channel conditions, the process of detecting the presence of the $P$ \acp{RAA} by a \ac{MIMO TRX} is equivalent to determining the top eigenvectors of the $P$ channels $\boldA^{(p)}$, $p=1,2, \ldots , P$, between the \ac{MIMO TRX} and the $P$ \acp{RAA}. In particular, \eqref{eq:bykbis} becomes
\begin{equation} \label{eq:bykbismulti}
\boldy[k]=\sum_{p=1}^P \left (e^{-\jmath \phi^{(p)}[k]} \, \boldA^{(p)}\left (\varphi^{(p)} \right ) \right )^*  \boldx^*[k-1] + \boldn^*[k]
\end{equation}  
where $\{ \phi^{(p)}[k] \}$ and $\varphi^{(p)}$ are, respectively, the information data and the \ac{AoA} associated with the $p$-th \ac{RAA}. 

If the \acp{RAA} are positioned at distinct angles and a favourable propagation condition is achieved by employing massive arrays (massive \ac{MIMO}), then the channels $\boldA^{(p)}$ become orthogonal \cite{SanguinettiBook:2017}. Therefore, their top eigenvectors can be obtained by estimating the $P$ top eigenvectors of the matrix ${\boldA=\sum_{p=1}^P \boldA^{(p)}\left (\varphi^{(p)} \right )}$.

In this regard, the iterative scheme proposed in Sec.~\ref{sec:algorithm}, which is limited to the estimation of only the top eigenvector, can be extended to estimate the $P$ top eigenvectors of $\boldA$, thereby enabling the detection of the $P$ \ac{RAA}-based devices along with the \acp{AoA} $\varphi^{(p)}$ of their signals. Specifically, once the top eigenvector has been detected, the detection of the second top eigenvector can be achieved using the same scheme described in Sec.~\ref{sec:algorithm}, provided that the iterative search is conducted in a space orthogonal to that spanned by the top eigenvector. To elaborate further, let's consider a matrix $\boldB$ that collects all the previously discovered top eigenvectors. At Step 4 of the scheme described in Sec.~\ref{sec:algorithm}, the following operation is performed:
\begin{equation} \label{eq:othogonal}
    \boldx[k]
    =\frac{\left ( \boldI - \boldB \, \boldB^{\dag} \right )  \boldy^*[k] }{\left \|  \left ( \boldI - \boldB \, \boldB^{\dag} \right )  \boldy^*[k]   \right \|} 
\end{equation} 
so that, before further processing, the updated beamforming vector $\boldx[k]$ is adjusted to be orthogonal to $\boldB$. This ensures that the subsequent search is conducted within the null space of $\boldB$, preventing the detection of previously identified eigenvectors (i.e., \ac{RAA}-based devices already detected).

In a more general scenario where the favourable propagation condition is not achieved, the top eigenvectors of $\boldA$ may not precisely align with the top eigenvectors of $\boldA^{(p)} \left (\varphi^{(p)} \right)$. This can result in interference among \ac{RAA}-based devices and subsequent performance degradation, as is typical in multi-user \ac{MIMO} systems.

\section{Convergence Analysis}
\label{Sec:Performance_Evaluation}
In the following, we analyze the convergence of the proposed iterative scheme to the optimum beamforming vector in the presence of data modulation and noise, by showing the time evolution of the \ac{SNR} at the \ac{MIMO TRX}. Then, we consider a dynamic scenario with the \ac{RAA} moving along a certain trajectory, and we discuss a channel tracking strategy to speed up the convergence and the performance. For the sake of simplicity in notation, we consider only one RAA-based node, although the same results apply in scenarios with multiple RAAs, provided that the channels are orthogonal due to favourable propagation conditions.

For convenience, let us introduce the eigenvalue decomposition of matrix~$\boldA$, which we consider now a generic modified round-trip channel matrix, as 
\begin{equation}
\boldA=\boldV \boldLambda \boldV^{\dag}=\sum_{j=1}^{N} \lambda_j \, \boldv_j \, \boldv_j^{\dag} \, 
\end{equation}
where $\boldLambda=\diag{\lambda_1, \lambda_2, \ldots, \lambda_N}$, being  $\lambda_j$ the $j$-th eigenvalue  with $\lambda_1\ge \lambda_2 \ge \ldots, \ge \lambda_N$, and $\boldv_j$ is the $j$-th eigenvector (direction) forming the $j$-th column of matrix $\boldV \in \mathbb{C}^{N \times N}$. 
As a consequence, the generic vector $\boldx[k]$ at the $k$-th iteration can be decomposed as
\begin{equation}
\boldx[k]=\sum_{j=1}^{N} x_j[k] \, \boldv_j 
\end{equation}
being $x_j[k]=\boldv_j^{\dag}\, \boldx[k]$ the projection of $\boldx[k]$ onto the $j$-th direction $\boldv_j$.
Similarly, the noise term can be expressed as
\begin{equation}
\boldn[k]=\sum_{j=1}^{N} n_j[k] \, \boldv_j 
\end{equation}
where $n_j \sim {\mathcal{CN}}\left (0, \sigma_j^2 \right )$, with 
\begin{equation}
\sigma^2_j=\sigma_w^2 + \frac{\lambda_j\, g\, \sigma^2_{\eta}}{\sqrt{\Ptx}}\, .
\end{equation}

 In the following analysis, we assume that the noise generated by the \ac{RAA}, which is transmitted back towards the \ac{MIMO TRX},  is negligible at the receivers' side compared to its own noise, i.e., $\sigma_j^2 \simeq \sigma_w^2$, $\forall j$. This is reasonable considering it is attenuated by the \ac{MIMO TRX}-\ac{RAA} channel. 

\subsection{Convergence and SNR Evolution}
\label{Sec:SNRevolution}
We now assess the \ac{SNR} in the $k$-th iteration of the scheme aimed at estimating the top eigenvector of the channel. This \ac{SNR} affects both the demodulation of the data symbol conveyed by $\phi[k]$ and the estimation of the angle $\varphi[k]$.

As per Step 4 of the scheme described in Sec.~\ref{sec:algorithm}, the beamforming vector $\boldx[k]$ at the $k$-th iteration is given by:
\begin{equation}
    \boldx[k]\!=\!\frac{\boldy^*[k]}{\left\|\boldy[k]\right\|}\!=\!\frac{\boldA \, e^{-\jmath \phi[k]} \boldx[k-1] + \boldn[k]}{ \left\| \boldA \, e^{-\jmath \phi[k]} \boldx[k-1] + \boldn[k] \right\|} 
\end{equation}
where
\begin{equation}\label{eq:ywithnoise}
    \boldy^*[k]=\sum_{j=1}^N \boldv_j \left ( \lambda_j x_j[k-1] e^{-\jmath \phi[k]}  + n_j[k] \right ) \, .
\end{equation}
According to step 7 in  Sec.~\ref{sec:algorithm}, the decision variable $u[k]$ at the $k$-th symbol  is 
\begin{align} \label{eq:uk1}
    u[k]&=  \boldx^{\dag}[k-1] \, \boldy^*[k] \\
    &= \boldx^{\dag}[k-1] \, \boldA \, e^{-\jmath \phi[k]} \boldx[k-1] + \boldx^{\dag}[k-1] \, \boldn[k] \nonumber \\
    &=e^{-\jmath \phi[k]} \, \sum_{j=1}^N \lambda_j \left|x_j[k-1]\right|^2  + \boldx^{\dag}[k-1] \, \boldn[k]  \nonumber
\end{align}
in which the first term is the useful one, as it contains the phase $\phi[k]$, and the second term represents the noise. 

Considering that by construction $\left\|\boldx[k]\right\|^2=1$, the SNR in  \eqref{eq:uk1} at the $k$-th time interval  is 
\begin{align}  \label{eq:SNRdec1}
\SNRdec[k]=\frac{\left ( \sum_{j=1}^N \lambda_j \, \left|x_j[k-1]\right|^2 \right )^2}{\sigma_w^2} \, .
\end{align}
Note that $\left |x_j[k-1]\right |^2 / \left \| \boldx[k-1] \right \|^2=|x_j[k-1]|^2$ represents the fraction of the total power transmitted by the \ac{MIMO TRX} associated with direction $\boldv_j$ at the discrete time $k-1$. Then, at the end of the $k$-th time interval, the  \ac{SNR} at the \ac{MIMO TRX} along the direction $\boldv_j$ is given by
 \begin{align}
 \mathsf{SNR}_j[k]=\frac{\lambda_j^2 \, |x_j[k-1]|^2}{\sigma_w^2} \, .
 \label{eq:SNRj2}
 \end{align}
Therefore, we can rewrite \eqref{eq:SNRdec1} as a function of $\mathsf{SNR}_j[k]$ as
\begin{align} \label{eq:SNRdecj}
\SNRdec[k]= \left ( \sum_{j=1}^N \frac{\mathsf{SNR}_j[k]}{\sqrt{\SNRjmax}}  \right )^2\end{align}
where 
\begin{align}
 \SNRjmax=\frac{\lambda_j^2 }{\sigma_w^2} 
 \label{eq:SNRj}
 \end{align}
 represents the maximum possible \ac{SNR} along the direction $\boldv_j$, i.e., the \ac{SNR} the receiver would experience if all the power were concentrated in the direction $\boldv_j$.

The goal is to determine an iterative expression for $\SNRdec[k]$ and evaluate the convergence condition of the iterative scheme proposed.
Considering \eqref{eq:ywithnoise}, the fraction of the total power that is associated with direction $\boldv_j$ at the beginning of time interval $k$  can be written as
\begin{align}
|x_j[k]|^2=\frac{\lambda_j^2 \, |x_j[k-1]|^2 + \sigma_w^2}{\sum_{i=1}^N \left ( \lambda_i^2 \, |x_i[k-1]|^2  + \sigma_w^2 \right )} \, .
\label{eq:etaj_def}
\end{align}
Then, by inverting \eqref{eq:SNRj2} and plugging $|x_j[k]|^2$ at both the left-hand and right-hand sides of \eqref{eq:etaj_def}, we obtain the following iterative formula for $\mathsf{SNR}_j[k]$ 
\begin{align}
\mathsf{SNR}_j[k]& =  \frac{\lambda_j^2\left(\mathsf{SNR}_j[k-1]+1\right)}{\sigma_w^2\left[\sum_{i=1}^N  \left ( \mathsf{SNR}_i[k-1] +1\right )\right]} \nonumber \\
&= \SNRjmax  \frac{\mathsf{SNR}_j[k-1]+1}{N+ \sum_{i=1}^N \mathsf{SNR}_i[k-1]}
\label{eq:SNRj1}
\end{align}
for $k \ge 2$, where 
\begin{equation}
\mathsf{SNR}_j[1]=\SNRjmax \left |x_j[0]\right |^2\, .
\end{equation}

The recursive expression \eqref{eq:SNRj1} can be numerically evaluated to obtain the \ac{SNR} evolution for each direction. Therefore, it is of interest to investigate whether convergence towards the top eigenvector is guaranteed, and under what conditions. Denote by $r=\rank{\boldA}$ the rank of matrix $\boldA$, i.e., the rank of the channel. 
Unfortunately, a general convergence analysis appears prohibitive for a channel with a generic rank. As a consequence, we derive in the following a convergence condition valid for a rank-1 channel and show numerically that the same result holds also for higher-rank channels.   
Then, assuming an ideal free-space \ac{LOS} channel for \ac{AoA} estimation as in \eqref{eq:LOSchannel}, we have only $\lambda_1\neq 0$, thus \eqref{eq:SNRdecj} is given by
\begin{align} \label{eq:SNRdec}
\SNRdec[k]=\frac{\left(\mathsf{SNR}_1[k]\right)^2}{\SNRmax}   
\end{align}
with the \ac{SNR}  at the $k$-th time interval along direction $\boldv_1$ in \eqref{eq:SNRj1} expressed as
\begin{align}
\mathsf{SNR}_1[k]=\SNRmax \, \frac{\mathsf{SNR}_1[k-1]+1}{N+ \mathsf{SNR}_1[k-1]}  \, .
\label{eq:SNR1}
\end{align}

This case allows an easy evaluation of the convergence value. In fact, the solution at the equilibrium of the recursive expression in \eqref{eq:SNR1} can be found by imposing that ${\mathsf{SNR}_1[k]=\mathsf{SNR}_1[k-1]=x}$ and solving the following second-order equation
\begin{align} \label{eq:eq}
x=\SNRmax\, \frac{x+1}{x+N} \, .
\end{align}
By considering only the positive solution, the condition on the final convergence value is: 
\begin{itemize}
    \item If $\SNRmax/N\gg1$, at the convergence it is ${\mathsf{SNR}_1[k]\simeq \SNRmax}$ and hence, from \eqref{eq:SNRdec}, ${\SNRdec[k]\simeq \SNRmax}$, which takes the role of asymptotic \ac{SNR} corresponding to the optimum beamforming vector for a channel with $r=1$. 
    \item If $\SNRmax/N\ll1$, we still have convergence but with $\SNRdec[k] \ll 1$ thus unable of guaranteeing a suitable data demodulation and \ac{AoA} estimation performance. 
\end{itemize}
Due to the previous result, we define the value ${\SNRboot=\SNRmax/N}$ as \textit{bootstrap SNR}. 
It is worth noticing that the convergence is always achieved to a convergence value which depends on the condition on the bootstrap \ac{SNR} above,  but it does not depend on the initial value $\boldx[0]$. 

Assuming the MIMO transmitter and the \ac{RAA} are in paraxial \ac{LOS} configuration (i.e., parallel arrays with $\varphi=\psi=0$), the first eigenvalue of $\boldA$ is, according to \eqref{eq:lambda1}, $\lambda_1=\sqrt{\Ptx} g \sigma_1^2$ and the maximum and bootstrap \acp{SNR} become, respectively, 
\begin{align} \label{eq:SNR0}
    \SNRmax&=\frac{\Ptx\, g^2 \, N^2\, M^2\, \Gap^2 \, \Gscm^2 \,  \lambda^4}{ \sigma_w^2 \, \left ( 4 \pi \, d \right )^4} \\
    \SNRboot&=\frac{\Ptx\, g^2 \, N\, M^2\, \Gap^2 \, \Gscm^2 \,  \lambda^4}{ \sigma_w^2 \, \left ( 4 \pi \, d \right )^4} \, .
\end{align}
According to the last equations, increasing $N$ and $M$ is beneficial for improving the link budget but also for the bootstrap \ac{SNR}; however, $M$ has a higher impact than $N$ thus, in general, it is more convenient to increase the \ac{RAA} size to improve the performance.

\begin{figure}[!t]
\centering
\includegraphics[trim=2.8cm 9.5cm 2.8cm 9.5cm, width=\columnwidth]{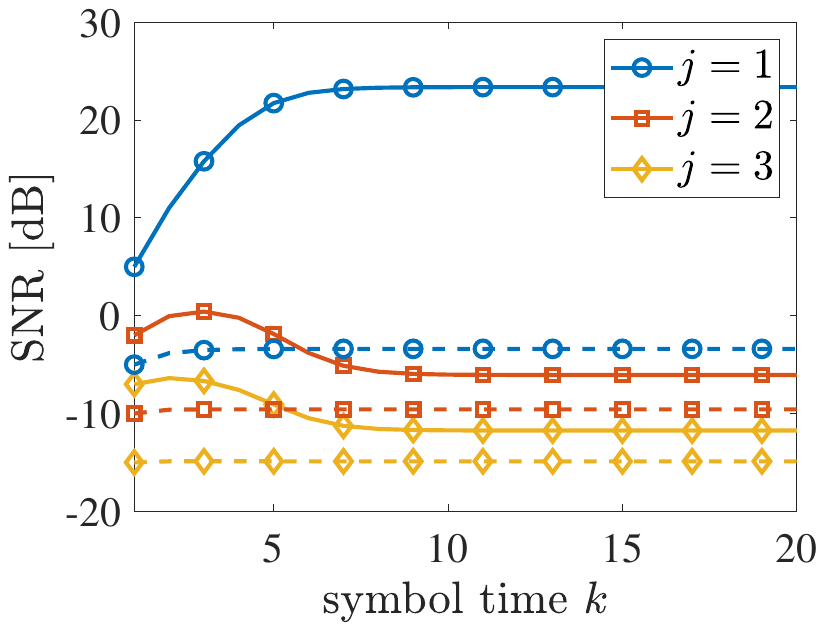}
\caption{Evolution of $\mathsf{SNR}_j[k]$, $j=1,2,3$ as a function of the time interval $k$ for a rank-3 channel with two different configurations. Continuous lines (--) are for \textit{Configuration a)} ($\SNRboot>\unit[0]{dB}$); dashed lines (-$\,$-) are for \textit{Configuration b)} ($\SNRboot<\unit[0]{dB}$).}
\label{Fig:SNRevolution} 
\end{figure}

\subsubsection{Numerical Example}
In order to show the convergence behavior analyzed above, we provide a numerical example for an array at the \ac{MIMO TRX} with $N=100$ elements  and a rank-3 channel with two different configurations: 
\begin{itemize}
\item \textit{Configuration a)} $\SNR_{1,\text{max}}=\unit[25]{dB}$, $\SNR_{2,\text{max}}=\unit[17]{dB}$, $\SNR_{3,\text{max}}=\unit[13]{dB}$, corresponding to  ${\SNRboot=\unit[5]{dB}}$; 
\item \textit{Configuration b)} $\SNR_{1,\text{max}}=\unit[15]{dB}$, $\SNR_{2,\text{max}}=\unit[10]{dB}$, $\SNR_{3,\text{max}}=\unit[5]{dB}$, corresponding to  ${\SNRboot=\unit[-5]{dB}}$.
\end{itemize}
The initial precoding vector $\boldx[0]$ is randomly chosen at the startup, therefore it is $\left |x_j[0]\right |^2 \simeq 1/N$ and ${\mathsf{SNR}_1[1] \simeq \SNRmax/N}$. Other initial strategies will be discussed in the next section.

In Fig. \ref{Fig:SNRevolution}, the evolution of  $\mathsf{SNR}_j[k]$, $j=1,2,3$, using \eqref{eq:SNRj1} is shown for the 2  configurations. 
From the plots, it can be noticed that when $\SNRboot>\unit[0]{dB}$ the \ac{SNR} associated with the top eigenvector ($j=1$)  converges to $\SNR_{1,\text{max}}$, given by \eqref{eq:SNR0}, within 5-6 time intervals, whereas it converges to very low values when $\SNRboot<\unit[0]{dB}$, as well predicted by the convergence condition even though it has been derived assuming a rank-1 channel. In fact, in any case the \ac{SNR} of the components of $\mathbf{x}[k]$ associated with the second and third eigenvectors, $\SNR_{2}[k]$ and $\SNR_{3}[k]$, tend to negligible values, i.e., the proposed scheme always converges to the top eigenvector but with a final \ac{SNR} depending on $\SNRboot$.  

According to the analysis outlined above, the convergence of the proposed scheme to $\SNR_{1,\text{max}}$ (i.e., to the largest SNR) is ensured only if the bootstrap \ac{SNR} significantly exceeds one. Therefore, the system must be properly designed, paying particular attention to factors such as the number of antennas, transmitted power, and RAA gain relative to the operating distance, to ensure that this condition is likely satisfied.

\subsection{Channel Tracking}
\label{Sec:Tracking}
\begin{figure}[t]	\centerline{\includegraphics[width=0.8\columnwidth]{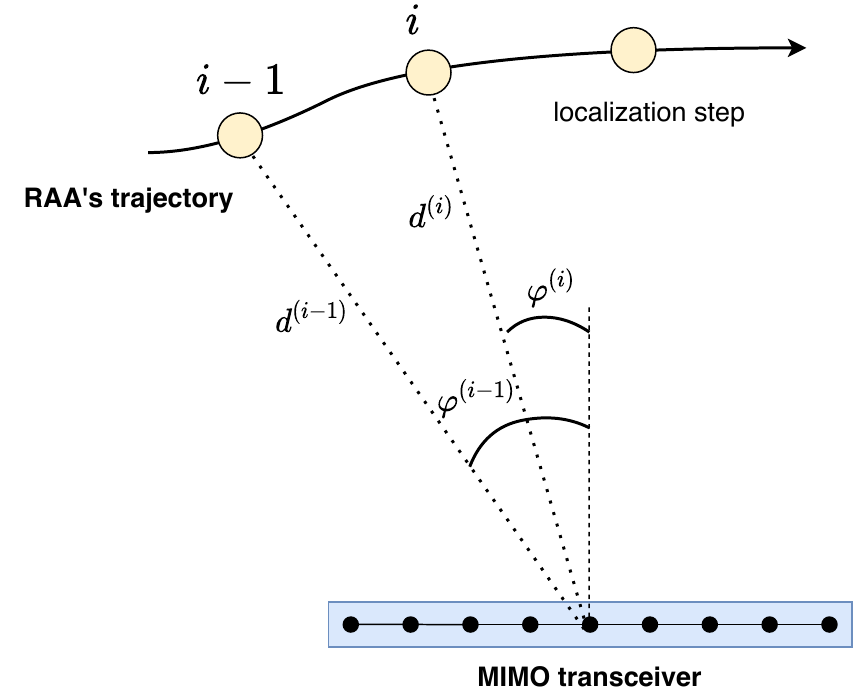}}
	\caption{Localization of the RAA in a dynamic scenario.}
	\label{fig:RAAtracking}
\end{figure}

Consider now the localization task in a dynamic scenario, where one of the two nodes (e.g., the \ac{RAA}) moves along a certain trajectory (see Fig.~\ref{fig:RAAtracking}). The iterative scheme described in Sec.~\ref{sec:AoARAA} is executed at specific points of the trajectory, namely, the localization steps. Specifically, at the $i$-th localization step an \ac{AoA} estimate (after \ac{RAA} detection) and a data packet consisting of $K$ symbols (e.g., the node's \ac{ID}) are obtained.
Referring to Fig.~\ref{fig:RAAtracking}, the \ac{RAA} is observed at an angle $\varphi^{(i)}$ with respect to the \ac{MIMO TRX} and located at a distance $d^{(i)}$ during the $i$-th localization step. Here, $\tau$ denotes the time interval between two consecutive localization steps. Thus, we can define the maximum localization update rate as $\mathcal{R}=1/\tau$, assuming that a sufficient number of angular measurements is collected at the $i$-th localization step to obtain an unambiguous location estimate.

To speed up the convergence of the proposed iterative scheme, rather than randomly choosing an initial beamforming vector $\boldx^{(i)}[0]$  at the $i$-th localization step, it may be more convenient to use the last available estimated beamforming vector $\boldx^{(i-1)}[K]$ at the previous localization step (i.e., $i-1$), thus assuming $\boldx^{(i)}[0]=\boldx^{(i-1)}[K]$. Since, according to this choice, the estimated beamforming vectors are re-used and updated iteratively during the movement of the node, we define such a strategy as \textit{channel tracking}. In the following, we will evaluate the conditions under which performing channel tracking is advantageous.

Let us now define, for further convenience, the \ac{SNR} along the first direction at the startup for the $i$-th localization step; this is the key parameter determining the convergence speed of the scheme. Specifically, we define it as \textit{startup SNR} so that $\SNRstart^{(i)}=\mathsf{SNR}^{(i)}_1[0]$ for the localization step $i$. We assume that at the localization step $i-1$ the convergence to $\SNR_{1,\text{max}}$ was achieved. 
Let us also now consider that the last beamforming vector $\boldx^{(i-1)}[K]$ of localization step $i-1$ is used as first beamforming vector $\boldx^{(i)}[0]$ of localization step $i$. In such a case, the \ac{SNR} along the first direction at the startup for the  localization step $i$, which determines the convergence speed, can be written as
\begin{equation}
\SNRstart^{(i)}=\gamma\,\SNRmax^{(i-1)} 
\end{equation}
where $\gamma$ reflects the change in the \ac{SNR} between the two positions when adopting the previous beamforming vector. In particular, we can decompose $\gamma$ as the product of two factors, i.e., $\gamma=\xi\rho$; the coefficient $\rho\le 1$ indicates the correlation between the channels related to the localization steps $i-1$ and $i$, while $\xi\lesseqgtr1$ indicates the difference in terms of path loss. 
The term $\rho$ can be obtained as the cross-correlation coefficient between the beamforming vectors corresponding to angles $\varphi^{(i-1)}$ and $\varphi^{(i)}$, that are, $\boldv(\varphi^{(i-1)})$ and $\boldv(\varphi^{(i)})$ as
\begin{equation}
\rho=\left|<\boldv(\varphi^{(i)}),\boldv(\varphi^{(i-1)})>\right|\, .
\end{equation}
The term $\xi$ can be written as the ratio between the path loss at positions $i-1$ and $i$, thus we have ${\xi=\mathcal{L}{(d^{(i-1)})}/\mathcal{L}{(d^{(i)})}}$ according to \eqref{eq:pathloss} when considering a free space scenario.
If $\SNRmax^{(i)}/N\gg1$ we have convergence at the maximum \ac{SNR} for the localization step $i$, which is ${\SNRmax^{(i)}=\xi \, \SNRmax^{(i-1)}}$.

The choice of using the previous convergence beamforming vector is beneficial only if the corresponding startup \ac{SNR} is larger than that obtained with the random guess $\boldx^{(i)}[0]$ at the $i$-th localization step. 
In fact, when selecting randomly the first beamforming vector $\boldx^{(i)}[0]$, we obtain the value  ${\SNRboot^{(i)}=\SNRmax^{(i)}/N}$ as the starting point of the iterative procedure. Formalizing, the choice is beneficial if
\begin{equation}\label{eq:ConvergenceCriterion}
\SNRstart^{(i)}=\xi\rho\,\SNRmax^{(i-1)}>\SNRboot^{(i)} = \frac{\SNRmax^{(i)}}{N}\, .
\end{equation}
However, since it holds that $\SNRmax^{(i)}=\xi\,\SNRmax^{(i-1)}$ (i.e., the \ac{SNR} obtained when considering the optimum beamforming vector at each location) the criterion \eqref{eq:ConvergenceCriterion}  becomes
\begin{equation}\label{eq:ConvergenceCriterion2}
\rho>\frac{1}{N}\, .
\end{equation}
It is very important to underline that the convenience is experienced in the number of iterations needed to reach convergence (i.e., convergence speed); no differences are obtained in terms of probability of convergence to the maximum SNR, since this is determined only by the bootstrap SNR value.

From \eqref{eq:ConvergenceCriterion2}, it could be inferred that a large number of antennas $N$ at the MIMO TRX is beneficial for ensuring faster convergence (i.e., that a larger number of antennas can allow to tolerate highly decorrelated channels between localization steps $i-1$ and $i$). However, it is important to note that the correlation coefficient $\rho$ itself strongly depends on the number of antennas $N$. 
For instance, consider a scenario where there is a small movement of the \ac{RAA} orthogonal to the \ac{MIMO TRX}'s array normal direction. In such a case, the primary source of change in the \ac{SNR} arises from the differing optimal combinations of phase values at the two positions, leading to $\gamma\approx\rho$.
Consider for simplicity $\varphi{(i-1)=0}$ (i.e., \ac{RAA} on the \ac{MIMO TRX}'s normal direction at localization step $i-1$), and the \ac{RAA} moving with constant speed $v$ transversal to the \ac{MIMO TRX}'s normal direction. When operating with half-wavelength spaced \acp{ULA}, we have 
\begin{align}
\boldv_1^{(i-1)}&=\frac{1}{\sqrt{N}} \left[1, 1, \ldots 1 \right]   \,\, \in \mathbb{C}^{N \times 1} \\
\boldv_1^{(i)}&=\frac{1}{\sqrt{N}} \left[1, e^{ \jmath {\pi}  \sin\varphi^{(i)}}, \ldots, e^{ \jmath {\pi} (N-1)  \sin\varphi^{(i)}} \right] \,\, \in \mathbb{C}^{N \times 1} \nonumber
\end{align}
thus we can write
\begin{equation}
\rho=\frac{1}{N}\left| \sum_{n=0}^{N-1} e^{- \jmath {\pi} n  \sin\varphi^{(i)}} \right| \approx \sinc{\frac{N  \sin\varphi^{(i)}}{2}}.
\end{equation}

In order to obtain a simple condition for determining the advantage of using the previous beamforming vector in the dynamic scenario, we can approximate the sinc function using its Taylor expansion around the origin, i.e., $\sinc{x}\approx{1-\frac{\pi^2 x^2}{3!}}$ and consider $\sin\varphi^{(i)}\approx\tan\varphi^{(i)}={v\tau}/{d^{(i)}}$ obtaining for \eqref{eq:ConvergenceCriterion2}
\begin{equation}\label{eq:ConvergenceCriterion3}
1-\frac{\pi^2 N^2  v^2 \tau^2}{24  \left(d^{(i)}\right)^2}>\frac{1}{N}\, .
\end{equation}
Then, by inverting \eqref{eq:ConvergenceCriterion3}, we get that the choice of using the previous beamforming vector is beneficial with respect to a random guess to ensure faster convergence only if
\begin{equation}\label{eq:ConvergenceCriterion4}
v<\frac{2 d^{(i)}\sqrt{6(N-1)}}{ \pi\tau   N\sqrt{N}}\, .
\end{equation}
When such a condition is not met, the channel variations between localization steps $i-1$ and $i$ are too significant, and relying on the previous beamforming vector proves ineffective. In such cases, a random guess can ensure faster convergence. 

According to \eqref{eq:ConvergenceCriterion4}, the larger the number of antennas $N$, the lower the maximum tolerated speed. In fact, a large number of antennas causes the channel to quickly decorrelate when moving from one position to another. Interestingly, it is worth noting that the convergence speed is not affected by $M$, i.e., the number of antennas at the \ac{RAA}. Thus, this parameter can be increased to improve the link budget without encountering convergence constraints in dynamic scenarios.

\subsubsection{Numerical Example}

\begin{figure}[t]
\centering
{\includegraphics[trim=2.8cm 9.5cm 2.8cm 9.5cm, width=\linewidth]{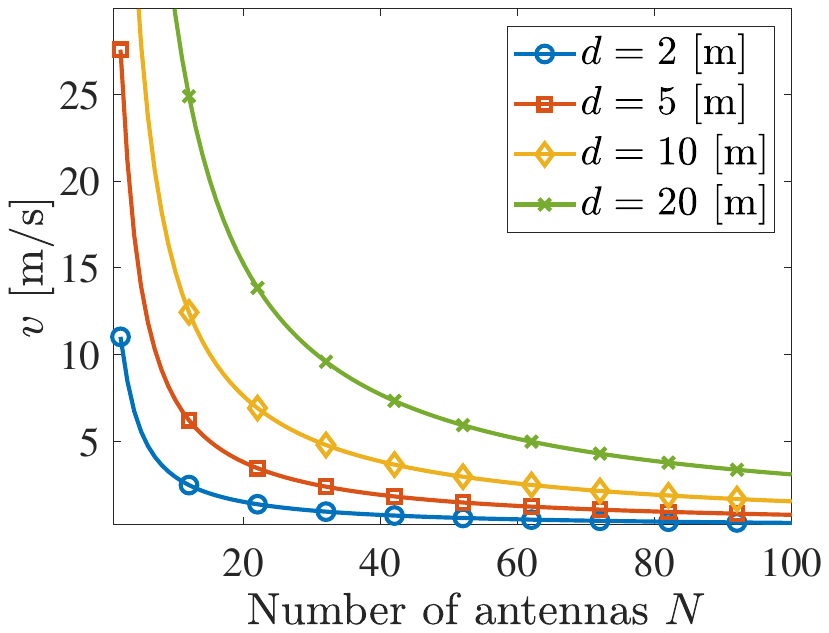}}
\caption{Maximum speed to make the channel tracking effective for improving the convergence speed of the proposed scheme.}
\label{fig:Convergence}
\end{figure}

In order to characterize the potential of tracking the channel by exploiting the last available beamforming vector, we consider a numerical example considering a variable number $N$ of antennas and $\tau=\unit[100]{ms}$. Fig.~\ref{fig:Convergence} shows the maximum speed for the \ac{RAA} in order to benefit from using the previous beamforming vector according to \eqref{eq:ConvergenceCriterion4}. It is possible to notice that, as the number of antennas increases, the maximum speed decreases. Moreover, if the \ac{RAA} is close to the \ac{MIMO TRX} the speed limit is lower since the channel changes faster its angular correlation (larger variation in angle $\varphi$ for a given transversal movement). It is worth noticing that if the speed is larger than that reported in Fig.~\ref{fig:Convergence}, convergence is still guaranteed if the bootstrap \ac{SNR} satisfies $\SNRstart^{(i)}/N\gg1$; however, using a random guess would be more effective (higher startup SNR thus faster convergence).

\section{Numerical Results}
\label{Sec:Numerical Results}

\begin{figure}
    \includegraphics[trim=2.8cm 9.5cm 2.8cm 9.5cm,  width=\linewidth]{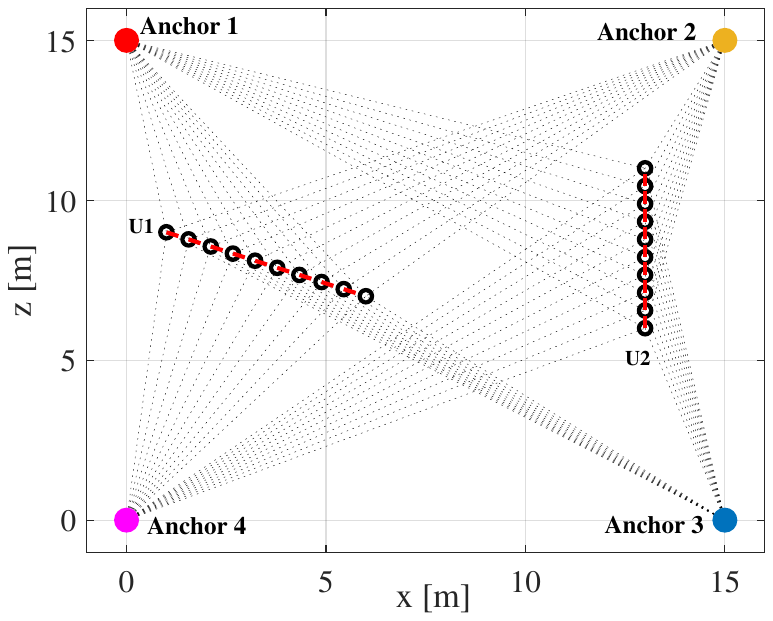}
    \caption{Simulated scenario: 2 users (U1 and U2) equipped with RAAs, moving along the red trajectories, 4 anchors equipped with MIMO TRXs. Black circles indicate the locations corresponding to the localization steps given a certain localization update rate $\mathcal{R}$.}
    \label{fig:SimulatedScenario}
\end{figure}

Since the basic building block and algorithm described in Sec. \ref{sec:AoARAA} adopted are common to both the envisioned architectures, we consider Architecture 1  as the reference for our numerical results, as described in Sec.~\ref{sec:schemes}. Simulations are performed in the 2D scenario depicted in Fig.~\ref{fig:SimulatedScenario}, where two \ac{RAA}-equipped \acp{UE} (U1 and U2) move along the red trajectories in the $x-z$ plane. Four anchors are also positioned in this scenario (\acp{MIMO TRX}).

The \acp{RAA} at the UE side consist of square antenna panels with $20\times20$ antenna elements arranged in the $x-y$ plane. Anchors are equipped with planar arrays featuring $10\times10$ antenna elements also deployed in the $x-y$ plane. Anchors employ the signals backscattered by the \acp{UE} to estimate their \acp{AoA}, according to the scheme outlined in Sec.~\ref{sec:algorithm}. The number of symbols (packet length) transmitted by the \acp{UE} is $K=40$.

A central frequency of $\unit[28]{GHz}$ is assumed, with ${W=\unit[10]{MHz}}$ bandwidth and $T=\unit[100]{ns}$ symbol time. The anchor antenna gain is $\unit[0]{dBi}$, and the noise figure is $\unit[3]{dB}$ for both the \acp{UE} and the anchors. A transmitted power of $P_\text{T}=\unit[0]{dBm}$ is assumed, with \ac{RAA} gains of $g=\unit[0]{dB}$ and $g=\unit[10]{dB}$ to represent passive and active \acp{RAA}, respectively. Threshold values of $\eta_1=\unit[30]{dB}$ and $\eta_2 = \unit[3]{dB}$ are selected for detecting a \ac{UE}. Consequently, depending on the channel condition and \ac{SNR}, one or more anchors in the scenario can detect the presence of the \acp{RAA} (i.e., \acp{UE}) at each localization step and obtain the associated \ac{AoA} estimates. Then, the \ac{AoA} estimates are fused using a least-square approach as described in \cite{PagVidBro:02}, yielding an estimate $\widehat{\mathbf{p}}=[\widehat{x}, \widehat{z}]$ of the position ${\mathbf{p}}=[{x}, {z}]$, in accordance with the geometry depicted in Fig.~\ref{fig:SimulatedScenario}.

\begin{figure}[t]
\centering
\includegraphics[trim=2.8cm 9.5cm 2.8cm 9.5cm, width=0.49\textwidth]{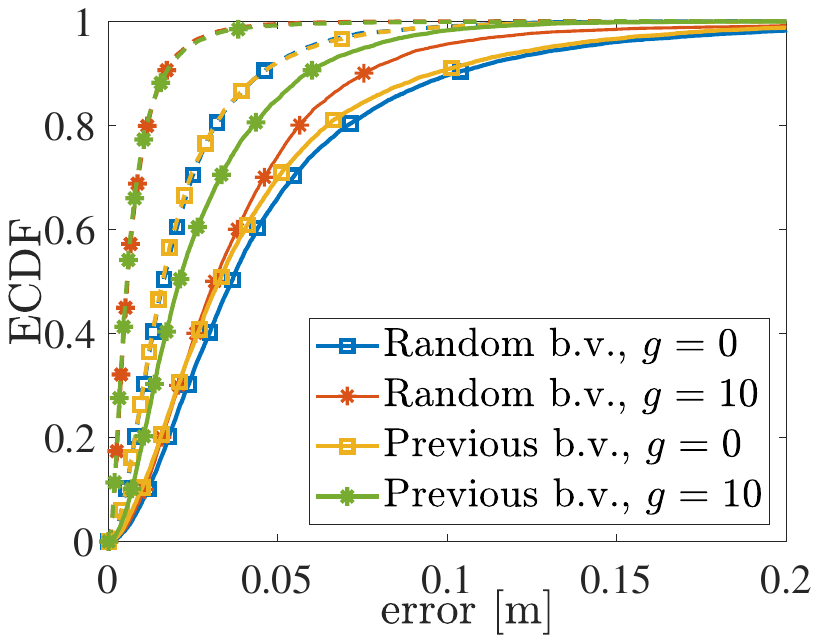}
\caption{Absolute position error. Dashed lines (-$\,$-) are for free-space LOS channels; continuous lines (--) are for 3GPP CDL-E LOS channels.}
\label{fig:erroreass}
\end{figure}

As a preliminary assessment of the proposed system performance, we conducted Monte Carlo simulations with a single \ac{UE}  (U1 in Fig.~\ref{fig:SimulatedScenario}). The results are reported in Fig.~\ref{fig:erroreass}, which shows the \ac{ECDF} of the absolute localization error $\epsilon=|\widehat{\mathbf{p}}-{\mathbf{p}}|$ under different conditions, obtained over 100 Monte Carlo iterations. Specifically, we considered a user moving with a speed of $v = \unit[0.54]{m/s}$ along a straight trajectory of length ${L= \unit[5.39]{m}}$, and a localization update rate $\mathcal{R} = \unit[10]{Hz}$ (i.e., $\tau=\unit[100]{ms}$), thus corresponding to  100 discrete localization steps along the trajectory. The results refer to various operating conditions, including:
\begin{itemize}
\item An ideal \ac{LOS} channel (free-space) and a realistic 3GPP CDL-E channel \cite{TR38.901:2019};
\item The exploitation of passive ($g=\unit[0]{dB}$) and active (${g=\unit[10]{dB}}$) RAAs;
\item The adoption of a random beamforming vector (b.v. in the plot legends) as initialization of the proposed scheme, or the last available beamforming vector, according to the channel tracking strategy presented in Sec.~\ref{Sec:Tracking}.
\end{itemize}

When focusing on the impact of the channel, it is evident from Fig.~\ref{fig:erroreass} that the best performance is achieved with free-space \ac{LOS} conditions (dashed lines), owing to the absence of multipath propagation. In this case, deviations in the \ac{AoA} estimate from the true angle primarily result from measurement noise. The performance experiences a slight degradation when considering the \ac{LOS} 3GPP channel (continuous lines), due to multipath effects. Generally, the localization error remains below $\unit[5]{cm}$  and $\unit[10]{cm}$  in $90\%$ of cases for the free-space channel (dashed lines) and CDL-E channel (continuous line), respectively.

The same figure also illustrates the impact of the path loss on performance. It is evident that leveraging active \acp{RAA} (red/green lines) reduces the localization error, primarily due to increased received power and consequently more robust \ac{AoA} estimation.
When considering the channel tracking mechanism proposed in Sec.~\ref{Sec:Tracking}, no differences in performance are observed for the free-space \ac{LOS} channel. This is expected, as channel tracking primarily facilitates faster convergence compared to randomly generating the initial beamforming vector. However, when a realistic multipath channel is considered, leveraging the previous beamforming vector (i.e., utilizing the channel tracking mechanism) results in a reduction in localization error. In fact, when the beamforming vector is randomly generated at each localization step, there is a chance that the \ac{AoA} estimator locks onto a multipath component, potentially leading to more significant errors with respect to starting the iterative algorithm from the previous, possibly correct, \ac{AoA} estimate.

\begin{figure}[t]
\centering
\vspace{1mm}
{\includegraphics[trim=2.8cm 9.5cm 2.8cm 9.5cm, width=0.49\textwidth]{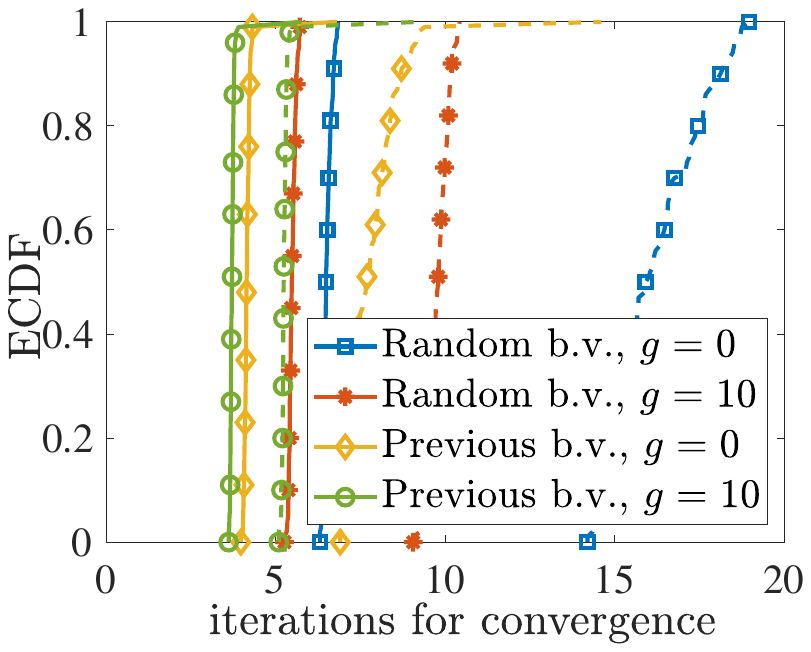}}
\caption{Number of iterations required for  detection relative to anchor 4, by varying the RAA gain and the use of the previous/random beamforming vector. Dashed lines (-$\,$-) are for U2; continuous lines (--) are U1.}
\label{fig:Speed_Convergence}
\end{figure}

The impact of the channel tracking mechanism, introduced in Sec.~\ref{Sec:Tracking}, is further investigated in Fig.~\ref{fig:Speed_Convergence}. This figure presents the \ac{ECDF} of the number of iterations required for \acp{UE}' detection relative to Anchor 4 (located at the bottom-left in Fig.~\ref{fig:SimulatedScenario}).
The \ac{LOS} free-space channel is here considered (similar outcomes are achieved with the 3GPP channel model). 
A noticeable difference in the number of iterations required for convergence is observed between using a random beamforming vector (blue/red lines) and the last available beamforming vector (yellow/green lines) as initialization of the proposed estimation scheme. 
The results demonstrate that utilizing the previous beamforming vector yields consistent improvement in terms of the number of iterations needed for convergence. This improvement is particularly significant for U2, which is the farthest from Anchor 4, thus experiencing highly correlated channels from one localization step to the other (as discussed in Sec.~\ref{Sec:Tracking}). Remarkably, when employing the previous beamforming vector, the convergence time is halved for $g=\unit[0]{dB}$. 

These results also offer insights into the localization update rate $\mathcal{R}$. While we set $\mathcal{R}=\unit[5]{Hz}$ in our simulations, much lower values could have been chosen, as the lower limit on this parameter is determined by the packet size $K$. In fact, the time elapsed between consecutive localization steps must be larger than the overall packet duration $KT$. Depending on the number of \acp{RAA} in the environment, to ensure the discrimination of each \ac{ID} a certain number of different \ac{PN} sequences must be available. By assuming, as an example, the adoption of M-sequences, a packet length of $K=1023$ symbols allows discriminating $60$ different \acp{RAA}, while a packet length of $K=8191$ symbols allows discriminating $630$ different \acp{RAA} \cite{DecGuiDar:J16}. These values lead to a maximum localization update rate $\mathcal{R}$ of roughly $\unit[10]{kHz}$ and $\unit[1]{kHz}$, respectively, which is larger than today's \acp{RTLS}, that usually provide tens of update per second. In fact, in this case, scalability is much simpler than in time-based \acp{RTLS}, where multiple users are usually interrogated sequentially. 
Since convergence is realized in a few iterations (e.g., $10$ iterations is a typical value according to Fig.~\ref{fig:Speed_Convergence}) the detection/estimation time results generally negligible with respect to the packet duration.

In addition to offering a very high localization update rate, the proposed solution offers several advantages over currently available solutions, such as \ac{UWB}-based \acp{RTLS}; in fact, it can work with narrowband transmissions, no synchronization is required as for time-difference-of-arrival based systems, and no clock drift is experienced. Moreover, being the \acp{RAA} backscattering devices, energy harvesting techniques can be included to make them energy autonomous.

\section{Conclusion} 
In this study, we have introduced two network architectures aimed at localizing mobile devices by harnessing the presence of backscattering \acfp{RAA}, which can be integrated either within the network infrastructure or directly onboard the mobile devices themselves. 
We have proposed an iterative scheme, which works on top of the two envisioned network architectures, which enables fast beamforming and \ac{AoA} estimation, without requiring any channel estimation procedure. Specifically, the full \ac{MIMO} gain is obtained, thus overcoming the main drawback of backscattering-based solutions caused by the two-way path loss, which can be detrimental when working at mmWave/THz. 
This results in simple, cost-effective, and energy-efficient devices, as no dedicated signaling or onboard signal processing capability for devices equipped with the \ac{RAA} is needed. Additionally, we have introduced an enhanced version of the proposed scheme capable of tracking the channel in the presence of mobile devices. This improvement yields even faster \ac{AoA} estimation when the mobile speed remains below a certain threshold, analytically characterized in this paper.

Numerical results have investigated both the localization accuracy and speed of convergence of the proposed scheme in free space channel conditions as well as when multipath affects the propagation. The results show that the proposed solution manages to keep the localization error at the centimeter level using narrowband signals and only a few anchor nodes while achieving a high localization update rate and very low latency, essential requirements in dynamic vehicular contexts.

\bibliographystyle{IEEEtran}
\bibliography{biblio_new}

\end{document}